\documentclass[usenatbib]{mn2e}
\newcommand{\hr}{\mbox{$^{\rm h}$}} \newcommand{\mn}{\mbox{$^{\rm
m}$}} \newcommand{\ssc}{\mbox{$^{\rm s}$}} \title{A multiband study of
Hercules A. II. Multifrequency VLA imaging} \author[Nectaria A. B.
Gizani and J. P. Leahy]{Nectaria A. B. Gizani,$^{1,2}$ and
J. P. Leahy,$^2$ \\ 
$^1$Centro de Astronomia e Astrof{\'{\i}}sica da
Universidade de Lisboa, Observat\'{o}rio Astron\'{o}mico de
Lisboa, Tapada da Ajuda, \\
$^{~}$1349-018 Lisboa, Portugal\\ 
$^2$University of Manchester, Jodrell Bank Observatory, Macclesfield, 
Cheshire SK11 9DL}

\begin{document}

\bibliographystyle{mn2e}
\maketitle

\begin{abstract}
We have mapped the powerful radio galaxy Hercules A at
six frequencies spanning 1295 to 8440 MHz using the VLA in all four
configurations. Here we discuss the structure revealed in total 
intensity, spectral index, polarization, and projected magnetic field.

Our observations clearly reveal the relation between the bright jets,
prominent rings, bulbous outer lobes and faint bridge that make up the radio
source. The jets and rings form a coherent structure with a dramatically 
flatter spectrum than the surrounding lobes and bridge, 
strongly suggesting that they represent a recently renewed outburst from 
the active nucleus. 
The spectrum of the lobes is also steeper than in typical radio sources, 
and steepens further towards the centre.
The compact core is optically thin and also
has a remarkably steep spectrum ($\alpha \simeq -1.2$).  There is some
evidence that the old lobe material has been swept up and
compressed ahead of the new outburst. We interpret the dramatic asymmetry
in the bright structure, and more subtle differences between diffuse
lobe structures, in terms of relativistic beaming combined with front-to-back
light-travel delays which mean that we view the two lobes at 
different stages of the outburst.

After correcting for Faraday rotation the projected magnetic
field closely follows the edge of the lobes, the jets, and the rings; 
the field pattern in the two lobes
is broadly similar. We confirm a strong asymmetry in depolarization 
and Faraday rotation, with the jet side the less depolarized and the
flatter spectrum, consistent with general correlations between these
asymmetries. The spectral index asymmetry is clearly present in the 
`old' lobe material and so, at least in this case, is not due to beaming;
but it can be understood in terms of the light-travel delay.

\end{abstract}

 
\section{Introduction}

Hercules A, is the fourth brightest DRAGN\footnote{For the definition
of the term DRAGN see \citet{Leahy1993} or Leahy, Bridle \& Strom
({\tt http://www.jb.man.ac.uk/atlas/}). It is slightly more general
than suggested by the acronym (Double Radiosource Associated with
Galactic Nucleus).} in the sky at low frequencies.  It is identified
with the central cD galaxy of a cluster at $z=0.154$, whose X-ray
emission was studied by \citet[hereafter Paper I]{GL2001}. The angular
size and width is $194 \times 70$~arcsec. For $H_0=65$ km s$^{-1}$
Mpc$^{-1}$ and $q_0=0$, which we assume throughout, 1 arcsec
corresponds to 2.8 kpc, giving a linear size and width of 540 and
$\simeq$ 200 kpc. The total radio luminosity is $\sim
3.8\times10^{37}$~W in the band 10~MHz to 100~GHz.

The radio jets and the galaxy major axis are aligned very well, with
position angles of 100\degr\ and $\sim 110\degr$ respectively.  This
alignment suggests that the galaxy is prolate in shape.
\citet{West1994} has cited this alignment as evidence for his model of
the formation of cD galaxies and powerful radio sources through highly
anisotropic mergers.

Optical observations with the HST \citep{Baum.etal1996}
through a broad band red filter discovered kpc scale rings of
obscuration, aligned near the radio axis with a slight offset from the
galaxy nucleus. The rings are $\approx 2$ arcsec in diameter and
the offset is $\simeq 1.5$ arcsec along the radio axis.
The host cD galaxy has a fainter optical companion
located $\simeq 4$ arcsec to the northwest. It is a typical faint
elliptical; as such its central surface brightness is actually much
higher than that of the cD.

The peculiar radio structure of Her~A was first revealed by 
\citet[hereafter DF84]{Dreher.etal1984}, using 
the VLA at up to 0.5 arcsec resolution.  It is an exception to 
Fanaroff \& Riley's (1974)\nocite{Fanaroff.etal1974} rule that DRAGNs with
$P_{\rm 178 MHz} > 1.5 \times 10^{25} {\rm \, W\,Hz^{-1}sr}^{-1}$ 
show `classical double' (FR\,II) structure.
Her~A has no compact hotspots, instead showing
an unusual jet-dominated morphology, with two jets which are quite different 
in appearance.  The western jet leads to a unique sequence of `rings' which
dominate the western lobe. 
The eastern jet is much brighter; indeed it has the highest flux density
of any jet found so far, and contributes around 40 per cent of the
total luminosity at 1.4 GHz.
Even allowing for the dependence of the FR division on host galaxy magnitude,
Her~A remains an extreme outlier, lying 30 times above the best fit 
transition line in the radio/optical luminosity diagram of 
\citet{LO96}.

In fact the radio structure of Her~A is not typical of FR\,I DRAGNs
either; in particular, its jets are well-collimated and knotty, more typical
of the `strong-flavour' jets in FR\,IIs than the `weak-flavour' jets
in FR\,Is \citep[c.f.][]{Bridle1992}.
For these reasons, Her~A is often classified as an intermediate case:
FR I/II. 

A wide variety of models, both kinematic \citep*{Mason.etal1988} and
dynamic \citep*{Meier.etal1991,SBS02} have been proposed 
to explain the formation of the jets and rings of Her~A.
Some \citep*[e.g.][]{Morrison.etal1996,SM02} are quite alien to the 
MHD-based twin-jet models normally applied to DRAGNs, and as yet
there is no consensus picture.

S. T. Garrington and G. Holmes (unpublished) have found a strong
depolarization asymmetry in Her~A, using low-resolution VLA
data. Because of its high flux density, Her~A is a perfect target to
study this effect at high resolution. Accordingly, in this paper we
present new, deep, VLA images in the 8-, 5- and 1.4-GHz bands.
The detailed structure revealed in total intensity, spectral index and
projected magnetic field casts new light on the unusual morphology.
Analysis of the Faraday rotation and depolarization is deferred to
Gizani, Leahy \& Garrington (in preparation; hereafter Paper III),
where it will be assessed in the context of the thermal gas
distribution studied in Paper I.

The rest of this paper is organized as follows. Sections~\ref{observe} \& 
\ref{reduction} describe our observations, data reduction and analysis. 
Results are
presented in Section~\ref{results}, where we present a detailed nomenclature
for the complex radio structure. Our discussion
begins with an analysis of the global 
depolarization and spectral index asymmetries (Section~\ref{alpha_prof}), 
followed by a discussion
of the geometric relation between radio-emitting and X-ray emitting plasma
in the cluster core region (Section~\ref{bridge}), and  of the 
collimation of the jets (Section~\ref{jetwidth}).  Section~\ref{beaming} 
considers what light is shed on the speed of the jets by the (lack of) 
symmetry between the jets in the two lobes. In Section~\ref{restarting}
we propose that the peculiar features of Her~A can be understood
if the jets have recently restarted. The implications of this idea for
the interpretation of the bright eastern jet and the rings are discussed
in Sections~\ref{disrupt} \& \ref{rings}. We summarise our conclusions
in Section~\ref{conclusion}.

\section{Observations}
\label{observe}
We have used the NRAO VLA, in the continuum mode with full
polarimetric imaging, to carry out our multi-band, multi-configuration
observations in the 8- and 1.4-GHz bands.  Table~\ref{obs} gives
the observational details.  The pointing centre is about 5 arcsec
south of the radio core, and was chosen to match earlier observations
at 5 GHz which we retrieved from the VLA archive and have
reprocessed.  These include some data published by DF84, together with
unpublished follow-up runs. In particular DF84 used an A-configuration
run with 50 MHz bandwidth and a pointing centre at the peak of the
eastern jet, which leads to appreciable bandwidth smearing at the core
and in the western lobe.  We use a later run with $2\times6.25$ MHz
bandwidth which used our common pointing centre, This gives negligible
smearing and allows a straightforward primary beam correction (see
below), although it has only half the sensitivity of the earlier
run. Calibrated 5-GHz D-configuration snapshot data was kindly
provided by Dr. S. Garrington.

\begin{table*}
\caption{VLA Observations of Her~A.}
\begin{center}

\begin{tabular}{ccccccl} \hline
RA & DEC & Config & Frequency & Bandwidth & Time &\multicolumn{1}{c}{Dates} \\ 
& & &  MHz & MHz & hrs & \\ \hline 
 16\hr 48\mn 40\fs 00 & +05\degr 04\arcmin 30\farcs 0  & C
        &  4885.1  & 50  & 7.5 & 1982-jan-10 \\
 16\hr 48\mn 40\fs 00 & +05\degr 04\arcmin 30\farcs 0  & A 
        & 4863.2/4813.2 & 6.25 & 6.5 & 1983-oct-24 \\
 16\hr 48\mn 40\fs 00 & +05\degr 04\arcmin 30\farcs 0  & B
        & 4872.6/4822.6 & 25  & 7 & 1984-feb-19 \\
 16\hr 48\mn 40\fs 10 & +05\degr 04\arcmin 28\farcs 0  & D 
	& 4885.1/4835.1 & 50  & 0.16 & 1989-nov-16 \\
& & A & L & 6.25 & 9 & 1995-jul-22 \\  
 16\hr 48\mn 40\fs 00 & +05\degr 04\arcmin 30\farcs 0  & B
 & X, L & 50, 12.5 & 9 & 1995-nov-12 \\ 
 & & C & X, L & 50, 12.5 & 4 & 1996-feb-23 \\ 
 & & D & X & 50 & 1.5 & 1996-aug-14 \\ \hline

\end{tabular}

{\footnotesize
Frequencies for the new observations: 8414.9/8464.9 MHz (X-band);
1664.9/1435.1 MHz and 1364.9/1295.0 MHz (L-band).}
\end{center}

\label{obs}
\end{table*}

Our new observations in A and B configurations were full tracks,
in order to obtain
adequate $uv$-coverage to produce an accurate image of the complex
structure of Her~A and to provide enough sensitivity at full
resolution. Approximately 70 per cent of the time was spent on the target
source, the remainder on calibrators and on driving.

We used 3C\,286 as the primary flux density and polarization angle
calibrator, and B1648+015 as the phase calibrator.
Because the source filled much of the primary beam at 8 GHz, 
in B, C, and D configurations we calibrated the antenna pointing 
each hour using B1648+015. We also pointed up on 3C\,286 before each flux
calibration scan.

For the 1.4-GHz band the time was split between two different pairs of
frequencies across the range 1295 to 1665 MHz, giving just over 3 hours per 
IF setting in A configuration. 
The frequencies used are listed in Table~\ref{obs}.

Because of the large size of Her~A and in order to avoid
bandwidth depolarization and radio-frequency interference (RFI), we
chose the narrow bandwidths of 6.25 and
12.5 MHz for the 1.4-GHz band. At 8 GHz 
we have made use of the full 50 MHz bandwidth and average the
two IFs together as there is unlikely to be significant Faraday
rotation between them. This bandwidth gives $\approx 10$ per cent beam
smearing at the outer edge of the source in full resolution maps; but as 
there is no compact structure there, this has little effect on the
images. In any case our main goal was to match the somewhat lower 
resolution of the 1.4-GHz data.

In B configuration the time was split 6:3
between the 8- and 1.4-GHz bands respectively, because the 1.4-GHz
observations were distributed to fill intermediate spacings.
In C-configuration only about 10 min was spent at each 1.4-GHz
IF setting.

The observations went almost as planned.  Some of the 1.4-GHz data
were affected by interference, and data were occasionally lost for other
reasons including lightning strikes in the A configuration.

\section{Data Reduction and analysis}
\label{reduction}
Data sets from each configuration  were separately
edited, calibrated and imaged in the standard NRAO
Astronomical Image Processing System (AIPS) software package.

\subsection{Mapping}
\label{self}

After initial external phase and amplitude calibration, the data 
were iteratively mapped and self-calibrated 
in the usual way \citep{Pearson.etal1984,Schwab1984}.  Her~A is a very 
strong source, so the self-calibration mechanism improved
dramatically the dynamic range of most maps.  At our highest resolution,
at 8 and 5 GHz, little flux is detected on the longest baselines and
to improve the signal to noise ratio 
in these cases we averaged right and left-hand
polarizations, and used averaging times of up to 15 min for amplitude
corrections.

In early cycles
the $uv$-range was restricted so that spacings underestimating the observed
visibilities were excluded. Two to four cycles of phase self-calibration 
were carried out and that initiated a new one with both amplitude and phase
solutions (two to four loops again). Each loop of deconvolving and
self-calibrating was carried out only if the noise was reduced
significantly.

The structure of Her~A, with bright compact (but resolved) features
embedded in diffuse emission, is difficult to image accurately, as calibration
errors and deconvolution artefacts associated with the bright features are
superimposed on the genuine diffuse emission. Furthermore, its position near
the equator means that the $uv$ tracks degenerate to nearly east-west
strips, causing sidelobes to build up in the north-south direction.
During cleaning, windows were used containing only
emission from Her~A as much as possible, restricting the area to
be cleaned only to the immediate vicinity of the source.
A straightforward
Clark CLEAN algorithm \citep{Clark1980} proved adequate down to a level
of $\sim 1$ mJy at 1.4 GHz, 
but after this point tended to develop strong CLEAN stripes.
Some improvement was obtained by using a `Prussian Hat' clean 
\citep{Cornwell1983}, which encourages smoothness in the clean
component model by adding a $\delta$-function to the centre of the beam. 

Since CLEAN stripes result from gaps in the $uv$-coverage, for deeper
imaging at 1.4 GHz we combined all the configurations and also the data
at 1365 MHz and 1435 MHz. We chose these data (rather than those at
1665 and/or 1295 MHz) because the combination gives excellent
$uv$-coverage on the longer baselines, and also because the data
suffered less from RFI and residual calibration errors. Further
self-calibration gave an A+B+C 1365, 1435 MHz map with
off-source noise $\simeq$ 0.1 mJy beam$^{-1}$, close to the
theoretical noise level.  This provided a set of clean components
which were used to self-calibrate separately the 1.4-GHz data at each
frequency.

To minimize striping in the final images 
the SDI method \citep*{Steer.etal1984} was used for the deconvolution.
We chose a resolution of 1.4 arcsec for the final images. Because of
the range in frequency in the 1.4-GHz band, this is smaller than the best-fit
beam at 1295 MHz (by about 14 per cent), and is lower than the full
resolution at 1665 MHz, allowing us to use robust weighting
\citep*{Briggs.etal1999} at that frequency.  The other 1.4-GHz band
frequencies were uniformly weighted.

In the same way as for the total intensity Stokes $I$ map, $Q$ and $U$
`dirty' maps were created and cleaned. Table~\ref{finma} shows
`cleaned' integrated flux, as well as integrated polarized intensity $p$
and off-source r.m.s. values of the final maps. The $Q$ and $U$ maps
are essentially noise-limited, while residual calibration errors
increase the fluctuations on the $I$ maps by factors of less than 2.
To maximize sensitivity in $I$, we also made a weighted average of the
4 single-frequency maps in the 1.4-GHz band, giving $\sigma$ = 0.095 mJy
beam$^{-1}$ at a nominal frequency of 1440~MHz.

\begin{table*}
\caption{Details of the maps}
\begin{minipage}{\linewidth}
\def\footnoterule{\kern-3pt
\hrule width 2truein height 0pt\kern3pt}
\begin{center}
 
\begin{tabular}{lllcrrrrrr} \hline
 
$\lambda$ & $\nu$ &  FWHM & Dynamic & $\sigma_{I}$ & $\sigma_{Q,U}$ & $\sum I$ & $\sum Q$ &  
$\sum U$ & $\sum p$ \\ 
cm & MHz &  arcsec & Range &  \multicolumn{2}{c}{mJy beam$^{-1}$} & Jy & Jy & Jy &
Jy \\ \hline 
3.6 & 8440 & 1.4  & 3200:1 & 0.023 & 0.013 &  5.968 &    0.416 & 0.37  & 1.49 \\
6 & 4848 & 1.4  & 4100:1 & 0.029 & 0.026 & 12.148 &    0.718 & 0.531 & 2.76 \\ 
18 & 1665 & 1.4  & 2100:1 & 0.132 & 0.073 & 41.158 & $-$0.220 & 0.827 & 5.12 \\ 
21 & 1435 & 1.4  & 2200:1 & 0.138 & 0.083 & 46.909 & $-$0.236 & 0.558 & 5.05 \\ 
22 & 1365 & 1.4  & 2700:1 & 0.121 & 0.085 & 48.658 & $-$0.237 & 0.542 & 5.09 \\ 
23 & 1295 & 1.4  & 2200:1 & 0.154 & 0.096 & 51.111 &    0.068 & 0.502 & 5.01 \\  \hline
3.6 & 8440 & 0.74 & 2400:1 & 0.011 & 0.011 &   ---  &     ---  &   --- & \\
6 & 4848 & 0.36 &  300:1 & 0.042 & ---   &   ---  &     ---  &   --- & \\ \hline
\end{tabular}
\end{center}
\end{minipage}

\label{finma}
\end{table*}

The two IFs at 8465 and 8415 MHz were combined and their mean value
(8440 MHz) is quoted in the tables. Because the 8415-MHz data were
affected by interference in D configuration, we initially
self-calibrated the C+D configuration data at just 8465 MHz, and used
the resulting model to self-calibrate the combined-frequency data.
These were then combined with the B-configuration data to give
uniformly-weighted maps at a full resolution of 0.74 arcsec.  Initial
maps from this process were used to self-calibrate the B-configuration
data only (as B-configuration phase artefacts initially prevent the
recovery of the faint, large-scale emission which dominates the
shorter baselines).  As the model improved the CD data were
self-calibrated as well, initially with long time constants just to
align positions and amplitudes between the datasets, after which all
data was processed together.  The fully self-calibrated data were
re-mapped with robust weighting and a 150 k$\lambda$ taper to give
maps at the standard 1.4-arcsec resolution.

In the archival 5-GHz data, different centre frequencies were used
in each configuration (see Table~\ref{obs}), but all data were
collected within a 100 MHz band. The observed Faraday rotation is too
low to cause differential rotation across the band, so we averaged all
the data, giving a weighted centre frequency of 5 GHz.  The old
data were affected by numerous phase glitches.  Affected data were
flagged, and the calibration/mapping cycle continued.  As at 8440 MHz,
we initially mapped the C+D data, then combined it with the B
configuration data.  Super-uniform weighting \citep{Briggs.etal1999}
was used to prevent the short-baseline data dominating the images; the
combined BCD data then gave a best-fit resolution close to our
standard 1.4 arcsec, and the final SDI clean image was restored with
this beam.

Correlator offsets in B-configuration produced
a $-0.8$~mJy artefact at the phase centre, 
just south of the core, and associated uncleanable
`sidelobes' to the north and south; fortunately these do not overlap
any detectable emission at 5 GHz.

The large images needed for the full A-configuration 
resolution at 5 GHz made it more efficient to use the
CLEAN+MEM method for imaging and self-calibration
described by \citet{LP91}.  
A particular problem for Her~A is that the inner jets are compact,
so best dealt with by CLEAN, but also very faint, so that to remove them
one would have to clean out all the flux in the source, which is what we
are trying to avoid by using MEM.  Our (partially-effective) solution was to
CLEAN a map from which the large-scale structure had been
filtered out by excluding the shortest baselines;
tight boxes were needed to avoid CLEANing the resulting negative 
sidelobes. The resulting CLEAN components were then subtracted from
the visibility data, which was then re-mapped using all baselines and
deconvolved using {\sc vtess}. 

The Maximum Entropy Method requires a `default image', usually chosen
to be a constant brightness at each pixel.  This choice causes flux to
spread into the off-source regions, giving large systematic errors on
short baselines. These errors can be suppressed by using a more
realistic default.  We found it essential to do so to get a model that
was usable for self-calibration. Specifically, we used an SDI-cleaned
image made from the 5-GHz BCD data.\footnote{Of course, the default
image must represent the same emission as the dirty map being
deconvolved; therefore in the CLEAN+MEM process, the CLEAN components
must be subtracted from the data before the default image is made.}

The signal-to-noise is low on the longest baselines, and so in A
configuration we only self-calibrated the amplitudes of the inner four
antennas on each arm.  The typical corrections found were $\la 2$ per
cent, and errors of this magnitude on the outer antennas will be
negligible, as they only contribute to the longest baselines.  
A tapered 4-configuration image at 1.4 arcsec resolution
was made but not used as it was slightly worse than the BCD image,
probably because of residual amplitude errors in A-configuration.

Because they were processed separately, the 8- and 5-GHz images
had to be accurately aligned with the 1.4-GHz A+B+C image, which was
done using the compact core.

The final total intensity maps were also corrected for the primary
beam attenuation of the VLA's 25-metre antennas \citep{green}, using
the AIPS task {\sc pbcor}.
At 1.4 GHz these corrections are negligible and were omitted.
At 8 and 5 GHz the corrections at the outer edge of the source
were 27 and 6.5 per cent respectively.

Although all our observations were calibrated against 3C\,286, this was
not strictly necessary as the integrated flux density of Her~A is known
from direct comparison to the absolute standards Cyg~A and Cas~A 
\citep{Baars.etal1977,Ott.etal1994}. We measured the total flux from
our primary beam corrected images, within a rectangular box just enclosing 
the source.  Our values agreed with the Ott et al.
spectral model to better than 1 per cent, except at 1435
and 1665 MHz, where the VLA fluxes are 1.6 and 3.9 per cent high, 
respectively. We have rescaled the images at these frequencies
to agree with the Ott et al. values.

\subsection{Error Analysis}

\label{error}

Until this point we have used the standard {\sc aips}
analysis to reduce our data.  Specialized software, described by
\citet*{Johnson.etal1995} was applied to the images to estimate
errors and propagate them into the final images of
depolarization $DP$, rotation measurement $RM$, spectral index
$\alpha$, and intrinsic magnetic field position angles.
For the error estimates we followed the prescription of 
Johnson et al. for VLA images with adequate sampling; 
that is, in quadrature with the off-source noise $n$ 
we added a fractional error
of $0.01S$ and an empirical term $0.15\sqrt{n S}$ where $S$ is the signal, 
and a term proportional to the image gradient corresponding
to r.m.s. misplacement of flux by 0.033 times the beamwidth. 

\subsection{Spectral index analysis}
\label{alpha_analysis}

We adopt the {\sc aips} sign convention for spectral index $\alpha$,
i.e. flux density $S_{\nu}\propto \nu^{\alpha}$. 

We calculated the spectral index at each pixel at 1.4~arcsec resolution
between 1.3 and 4.8 GHz, and between 4.8 and 8.4 GHz. In the former case
we used a weighted least squares straight line fit in the log--log plane
to the 5-frequency data. We required a signal-to-noise ratio of $\ge3$:1 at
all frequencies to include data at a given pixel in the fit (in practice
the limiting factor is the 4.8-GHz image). 

Her~A contains several regions where components with substantially
different spectral indices are superimposed on the same line of sight,
so directly-evaluated spectral indices give just a weighted average.
To get estimates for each component 
we used the so-called `spectral tomography' technique 
of \citet{Rudnick.etal1996} and \citet{Katz-Stone.etal1997}.
In this method, images at two frequencies are scaled and differenced
so that material with a particular spectral index is cancelled. The
process is repeated for a range of assumed $\alpha$ values. By comparing
the sequence of images produced (e.g. as a `movie' on a TV display), 
one can estimate the spectral index at which a particular feature vanishes,
in the sense of having negligible contrast with surrounding material.

\subsection{Polarization Analysis}
\label{pol_analysis}
In the presence of noise, the polarized intensity $p = \sqrt{Q^2 +
U^2}$ suffers a slight positive Ricean bias on
average. Therefore $p$ has been corrected for Ricean bias
\citep{Simmons.etal1985,Leahy3.etal1989} by subtracting the error at
each pixel in quadrature. $p$ was set to zero where the observed
polarized intensity was less than the error.  From the corrected $p$
images we calculated the fractional polarization $m = p/I$.

We have analysed the multi-wavelength data to derive the Faraday Rotation
measure and projected magnetic field ($B$-field) direction, using
the algorithm described by \citet{Johnson.etal1995}. 
In this paper we only present the results for the $B$-field, and so
defer full discussion to Paper III.  However, we note that in the
western lobe, depolarization is so strong that the $\lambda^2$-law
breaks down at 1.7 GHz, so the $RM$ must be determined between 5 and
8 GHz only.  This leads to ambiguities due to the unknown number of
half-turns, $n\pi$, between the two wavelengths. We have inserted these
by hand, guided by continuity in the $RM$ and $B$-field maps, together with
the depolarization maps which reveal genuine abrupt changes in
$RM$.

\subsection{Measuring the collimation of the jets}
\label{gausfit}

We tracked the width of the jets by measuring
$\theta_J$, the FWHM of 1-D Gaussian fits to
slices oriented at PA 10\fdg 1, approximately perpendicular to the
inner jets.  The jets curve slightly, so our slices
are not always precisely perpendicular, but the misalignment is
no more than 12\degr, giving a negligible over-estimate of the widths
($\la 2$ per cent).

We used our full-resolution maps at
both 5 and 8 GHz; in addition, as the signal-to-noise is low in the
inner jets at 5 GHz, we smoothed the 5-GHz image to give an elliptical 
beam with FWHM $1.08\times 0.36$ arcsec, with its major axis along the
jet. Between 35 arcsec east of the core and 16 arcsec west,
we fitted each profile with a single Gaussian and a linear baseline,
using {\sc xgauss}. 
Further from the core we had to model underlying diffuse
emission with
a second broad Gaussian, and this was done interactively using {\sc slfit}.
In all cases the FWHM beam width (perpendicular to the jet) was subtracted in
quadrature, to give an estimate of the deconvolved jet width.

As the jet profile is generally not Gaussian, $\theta_J$ is just
an empirical definition of `width'.
Note that the fitting behaves somewhat differently
in the case where the jet is well resolved, when the fit is controlled
by real residuals from Gaussian shape, and in the case where the jet
is barely resolved, when the profile is nearly Gaussian and our fit
effectively measures the second moment of the brightness distribution.
Comparison of the results from the 5- and 8-GHz maps
demonstrates this point: where the jet is well
resolved or poorly resolved in both maps, agreement is excellent; but
between 8 and 20 arcsec east of the core, the jet is well resolved at
5 GHz and not at 8 GHz, and the results are systematically different.

\section{Results} 
\label{results}

\subsection{Total Intensity}

\subsubsection{Overview}
\label{toti}
\begin{figure*}

\centering
\setlength{\unitlength}{1cm}
  
\begin{picture}(17.5,10.0) 
\put(0,10){\includegraphics{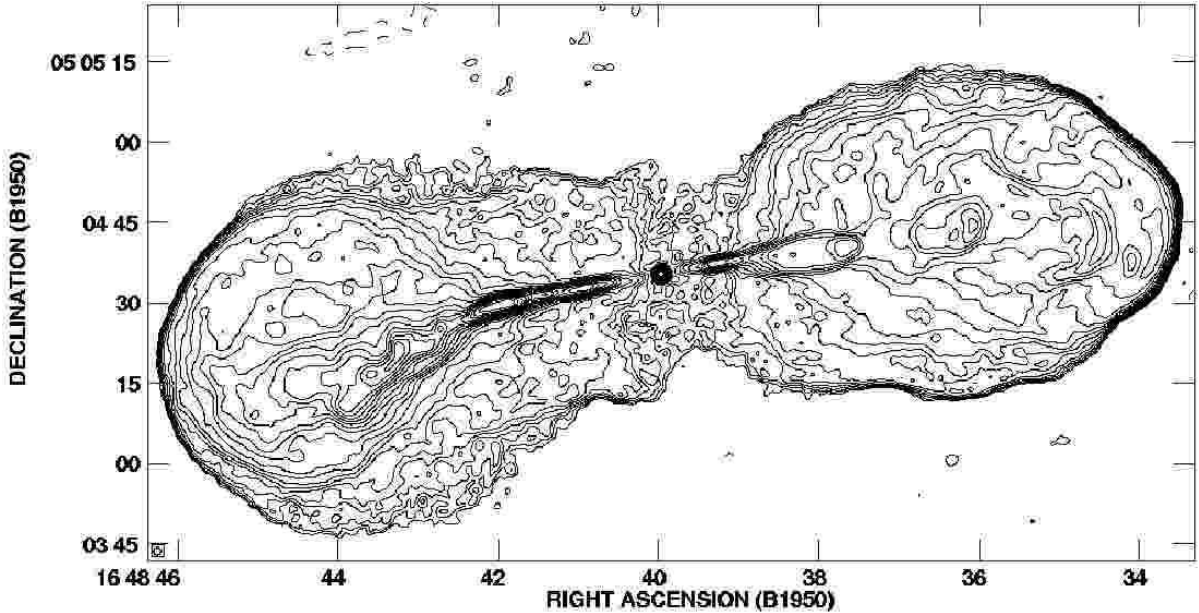}}
\end{picture}

\caption{A combined map of the total intensity distribution of Her~A
at the nominal frequency 1440 MHz.  The beam size (1.4 arcsec) is
shown in the lower left-hand corner.  Contours are logarithmic,
separated by factors $\sqrt{2}$, and starting at $\pm 0.3$ mJy
beam$^{-1}$.}
\label{20toti}
\end{figure*}

Fig.~\ref{20toti} presents contours of the total intensity of Her~A at
1440 MHz.  The source extends overall about 194 arcsec ($\simeq$ 540
kpc), and its maximum width is $\simeq 70$ arcsec ($\simeq$ 200 kpc).
The inner jets, although prominent with logarithmic contouring,
are actually quite faint (c.f. Fig.~\ref{xjets} below), but the eastern jet 
brightens dramatically in the outer lobe.
The two lobes are almost symmetric in outline: the western
extends about 97 arcsec (270 kpc) and the eastern $\approx 96$
arcsec (267 kpc).  The radio emission is generally bounded by a
sharply-defined perimeter at which the intensity drops by an order of
magnitude or more within one beamwidth. Exceptions to this are north
and south of the bright parts of the eastern jet, at 16\hr 48\mn
44\ssc\ to 41\fs 5, and the very faint bridge emission just west of
the core.  The lack of definition in the
former region may be due to residual sidelobes from the jet; the
deepest published map, that of \citet{Kassim.etal1993} at 330~MHz,
shows that we are missing little or no emission there; the western
bridge is slightly wider in the 330 MHz image than in ours, with a
full width of $\simeq 40$~arcsec ($\simeq$ 110 kpc), about the same as
that east of the core.

Although not obvious from our contour map, the south-western edge of
the western lobe is edge-brightened, consisting of long thin
filaments or possibly a single one. There are also signs of
edge-brightening in the eastern lobe but this may be confused by
residual artefacts parallel to the jet.

\begin{figure}
\centering
\setlength{\unitlength}{1cm}
  
\begin{picture}(8.5,4.5) 
\put(0.5,0){\includegraphics{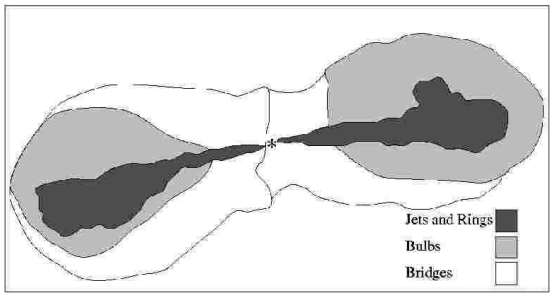}}

\end{picture}
\caption{Sketch of the main components of Her~A. An asterisk marks
the core. More detailed labelling of features in the jets and rings is
given in subsequent figures.}
\label{sketch}
\end{figure}

DF84 identified three distinct kinds of structure in Her~A, all
of which are visible in Fig.~\ref{20toti}, and sketched in
Fig.~\ref{sketch}.  The source is dominated by the high-brightness
features, namely the jet on the eastern side, and on the west the
narrow counter-jet, followed by the famous series of `rings'. The
surrounding lobe emission can be divided into two components: the
relatively bright outer part, hereafter the `bulb', and the much
fainter `bridges', which contain the long thin filaments noted
earlier.  Although the two bridges meet at the centre, there is a
noticeable brightness minimum between them and the eastern bridge is
several times brighter than the western, so they do seem to be two
distinct structures.

\subsubsection{The Bulbs}
\label{bulbs}

\begin{figure*}
\centering
\setlength{\unitlength}{1cm}
\begin{picture}(17.5,22)
\put(2,-0.5){\includegraphics{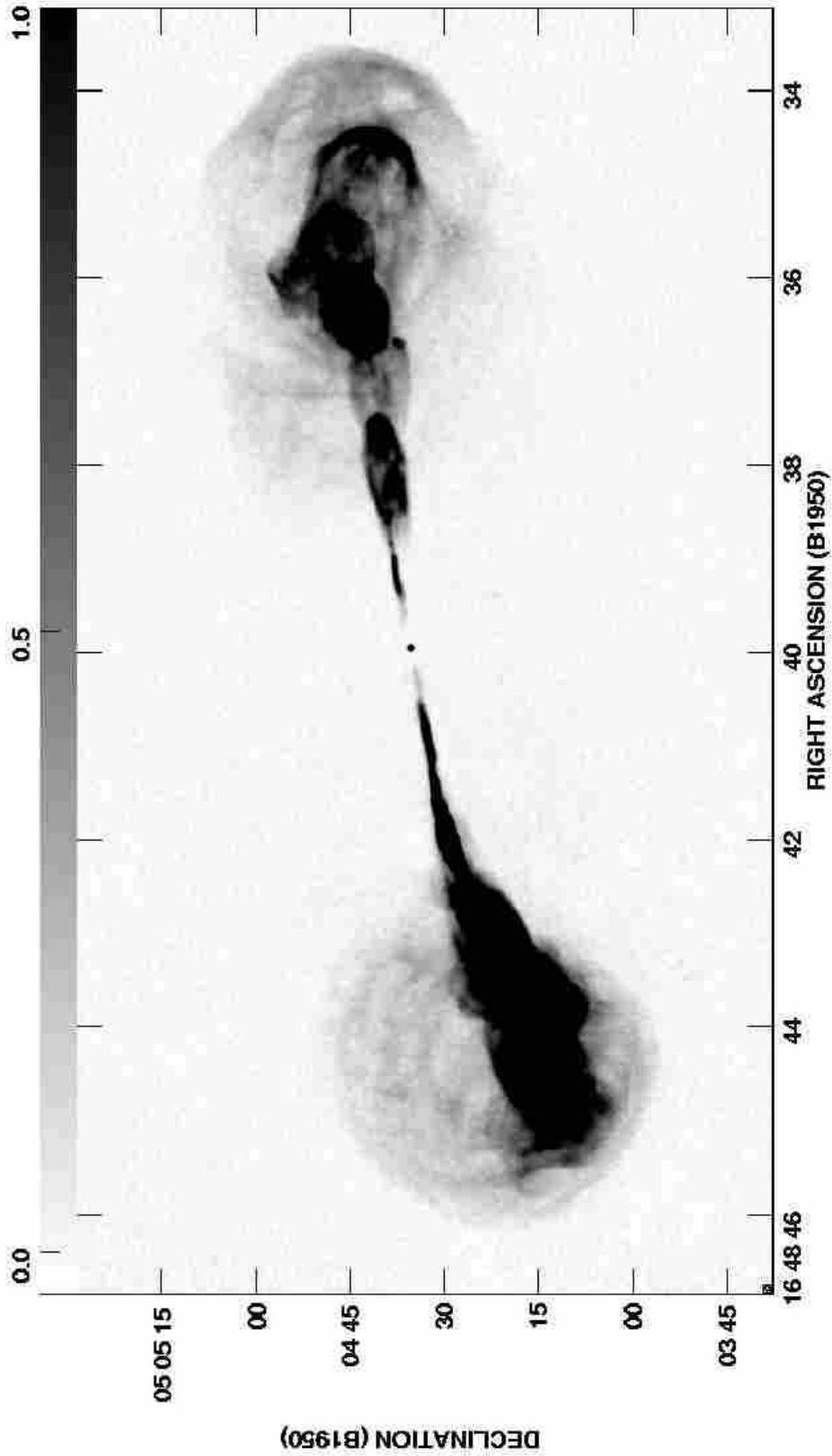}}
\end{picture}
\caption{Total intensity distribution of Her~A at 8440 MHz.
The beam size (0.74 arcsec) is shown in the lower left-hand corner. 
The grey-scale runs from -0.033 to 1 mJy.}
\label{xtoti}
\end{figure*}

Fig.~\ref{xtoti} shows a grey-scale of
the full-resolution image at 8440 MHz, with the high-brightness features
burnt out to reveal the structure of the bulbs. At this 
resolution and frequency the bridges are too faint to detect.

The two bulbs are notably different.
The eastern is brighter and nearly circular, while the western is more oval.
Both show arc-like filamentary structure but they differ in character.
The filaments in the east are thicker and clumpier, those in the west are
more wispy. As noted by DF84, the outer edge of the western lobe, though
arc-like, is different in character from the bright rings; it seems to 
contain several overlapping filaments and is continuous with the diffuse
lobe emission rather than the high-brightness features.

\subsubsection{The jets and rings}

\begin{figure*}
\centering
\setlength{\unitlength}{1cm}
\begin{picture}(17.5,11)
\put(-0.3,11){\includegraphics{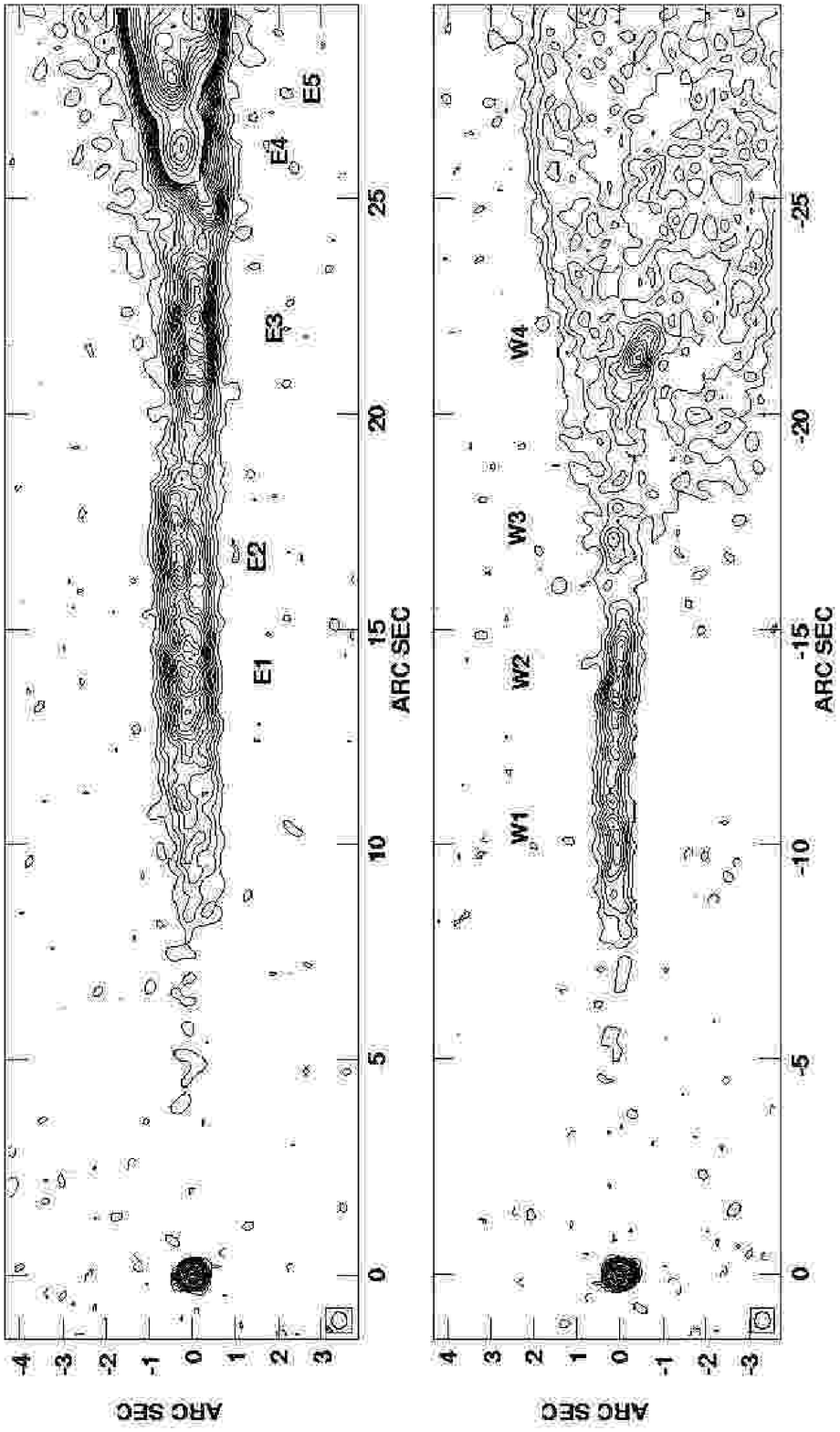}}

\end{picture}
\caption{The inner jets at 4848 MHz with 0.36 arcsec resolution. 
For ease of comparison the maps have been rotated 
by $-100\fdg 1$ (East; top), and  
$-10\fdg 1$ (West; bottom). 
Coordinates give distances from the core.  Contours are at 
$(-1, 1, 2, 3\ldots 16, 24, 32\ldots)\times 0.13$ mJy beam$^{-1}$.}
\label{ABCDjets}
\end{figure*}

\begin{figure*}
\centering
\setlength{\unitlength}{1cm}
\begin{picture}(17.5,10)
\put(0,11){\includegraphics{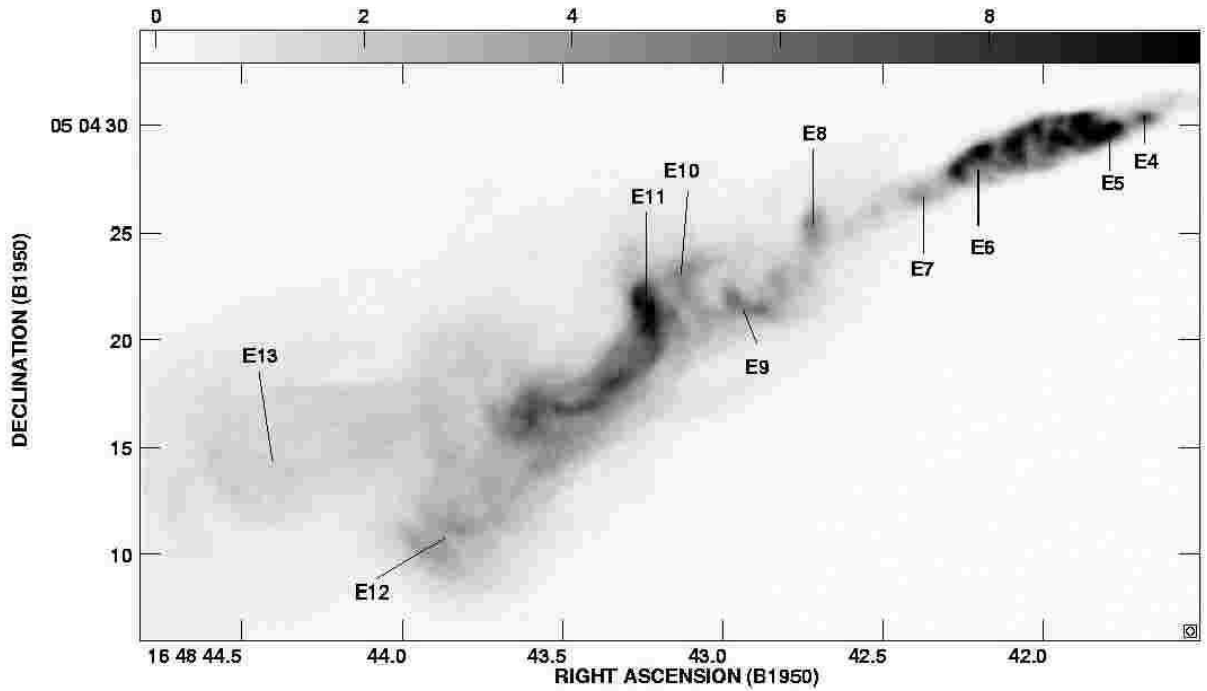}}
\end{picture}
\caption{The outer eastern jet at 4848 MHz with 0.36 arcsec
resolution. The grey-scale runs from $-0.13$ to 10 mJy beam$^{-1}$.}
\label{ABCDeast}
\end{figure*}

\begin{figure}
\centering
\setlength{\unitlength}{1cm}
\begin{picture}(8.5,5.8)
\put(-0.5,-0.5){\includegraphics{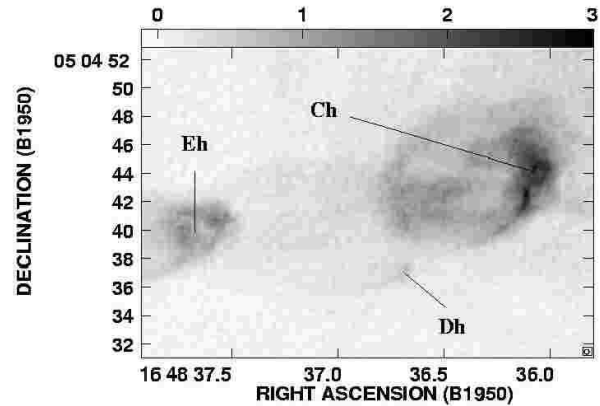}}
\end{picture}
\caption{The mini-ring and ring C at 4848 MHz with 0.36 arcsec
resolution. The grey-scale runs from $-0.1$ to 3 mJy beam$^{-1}$.}
\label{ABCDwest}
\end{figure}

The jets are first detectable $\approx$ 2.9 and 3.8 arcsec 
east and west of the core respectively. 
The initial 4 arcsec of each jet, best seen in Fig.~\ref{xtoti}, 
is extremely faint,
Fig.~\ref{ABCDjets} shows the inner jets in the full resolution 4848 MHz image,
while Figs~\ref{ABCDeast} and \ref{ABCDwest} show the bright features in the
eastern and western lobes respectively.
We have labelled knots in the western and eastern jets as W1--W4 and E1--E13.

At our highest resolution the eastern jet is moderately to well
resolved.  The centre of the brightness profile is usually rather
flat-topped although occasionally it can be quite sharply peaked; the
implication is that in three dimensions, the emission is low along the
jet axis and the jet has more emission near the surface
\citep[c.f.][]{Owen.etal1989b}. Beyond knot E11 the jet
contains (or may be composed of) a number of filaments running roughly
parallel to the overall axis, one of which curls around at the end of
knot E13.

At first sight the edges of the jet look well defined, but comparison
between the jet brightness profiles and that of a uniformly-filled
cylinder\footnote{This provides a reference model with a sharp outer
boundary. The crucial region near the edge of the jet is not affected
by the possible presence or absence of a dim region along the axis.}
convolved with a Gaussian beam showed that the edges of the jets are
less sharp, suggesting that there is a boundary layer in which the
intensity declines slowly to zero.

\begin{figure*}
\centering
\setlength{\unitlength}{1cm}
\begin{picture}(17.5,10.5)
\put(0.2,10){\includegraphics{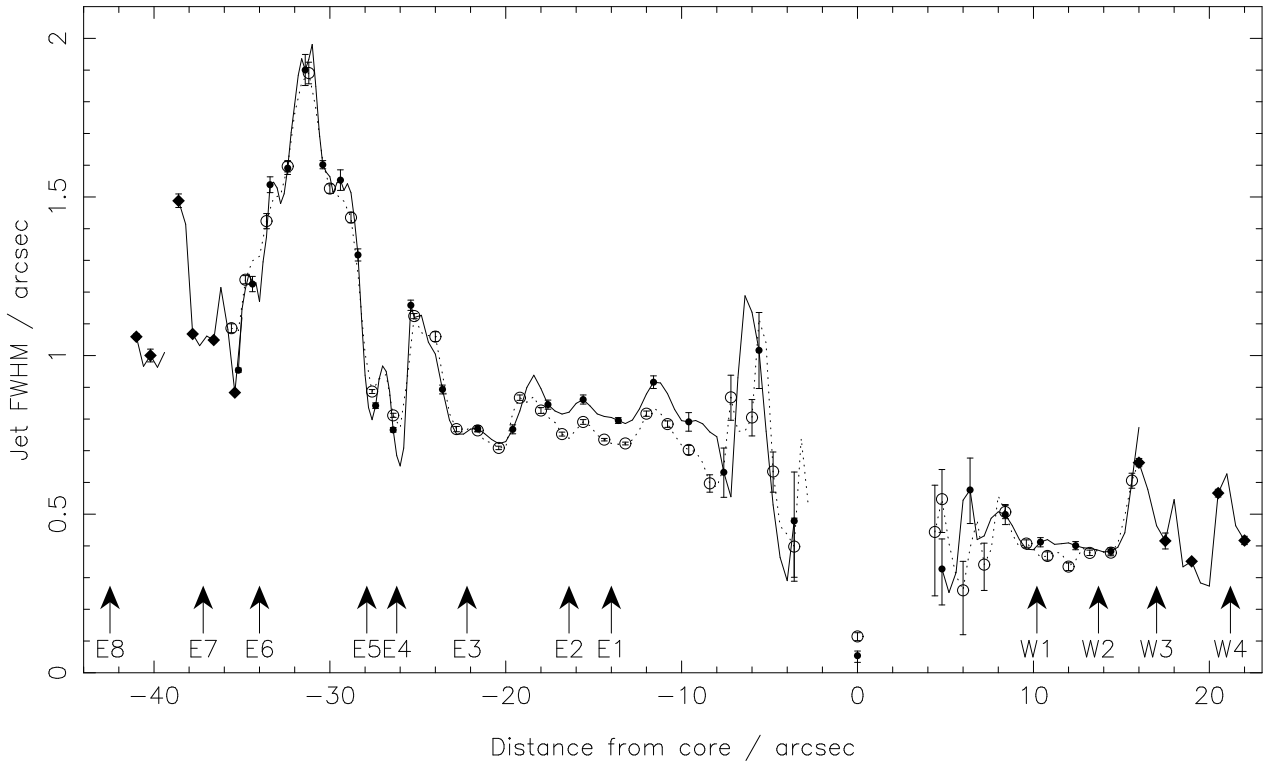}}
\end{picture}
\caption{Collimation of the Her~A jets, represented by the deconvolved
FWHM from 1-D Gaussian fits ($\theta_J$ in the text), plotted against
$l$, the distance from the core measured in PA $-$80\fdg 1.  Negative
distances are for the eastern jet. The positions marked for Ei and Wj
(for i=1 to 8 and j=1 to 4) are the brightness peaks of the knots. The
lines marked with circles are fits from {\sc xgauss}; the lines marked
with diamonds are from {\sc slfit}.  Solid line and filled points:
from the smoothed 4848 MHz map with FWHM beam $1.08\times0.36$ arcsec;
except east of $-20$ arcsec, where we used the full-resolution 4848
MHz map.  Dashed line, open points: from 8440 MHz map at 0.74 arcsec
resolution. Only a subset of the measured values are marked with
symbols and error bars, to avoid overcrowding. The error bars have
been corrected for oversampling \protect\citep[c.f.][]{Cond1997}.
These show how much the fit is uncertain due to noise; they do not
include systematic errors due to fitting a Gaussian to a non-Gaussian
profile.}
\label{collim}
\end{figure*}

Fig.~\ref{collim} plots the FWHM $\theta_J$ from Gaussian fits to slices 
across the jet against $l$, the distance from the core.  We measure the
eastern jet up to knot E8, after which it is too disrupted to allow
meaningful measurement of collimation. 
Roughly, the full width 
may be 1.5--2 times $\theta_J$. For comparison,
Gaussian fits to a top-hat and filled cylinder of width $w$ give
$w = 1.14 \theta_J$ and $1.75 \theta_J$ respectively. 

\begin{figure*}
\centering
\setlength{\unitlength}{1cm}
\begin{picture}(17.5,16)
\put(-0.5,0){\includegraphics{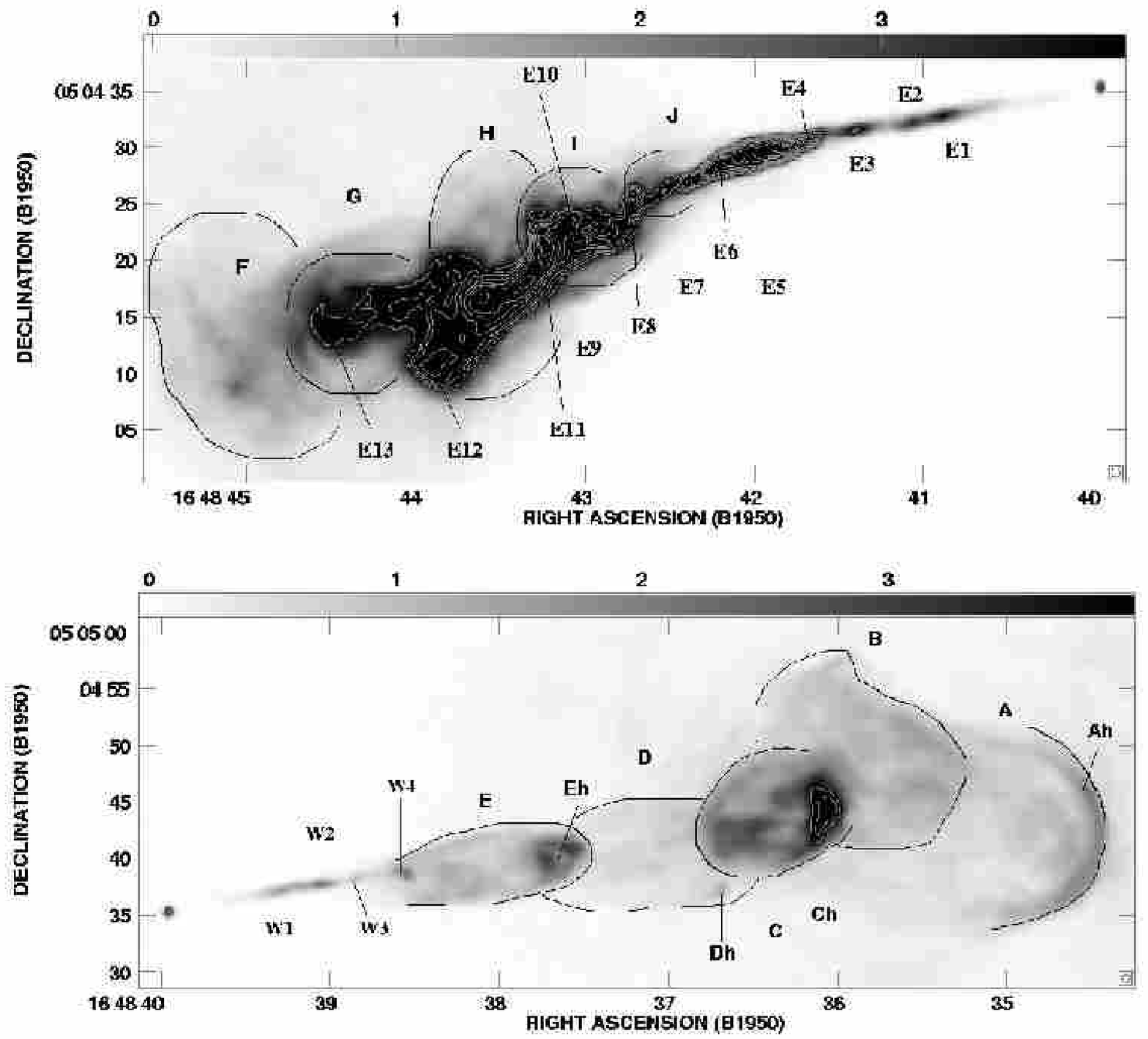}}

\end{picture}
\caption{Close-ups of the high-brightness structure at 8440 MHz.
Grey-scale runs from $-0.033$ to 4 mJy beam$^{-1}$ and contours are at
4, 5, 6, 8, 10, 12, 16, 20, 24 mJy beam$^{-1}$.  Top: eastern jet and
rings. Bottom: western jet and rings.  The angular scale is the same
in the two plots. We have labelled the ring features and jet knots
discussed in the text; for clearer identification of some of the
eastern knots, see Fig.~\ref{ABCDeast}.}

\label{xjets}
\end{figure*}

Structure at intermediate brightness is shown in Fig.~\ref{xjets},
which gives details from the 8440 MHz image with lower contrast than
Fig.~\ref{xtoti}.  The western jet is followed by the sequence of
rim-brightened features which DF84 christened `rings'. We retain this
word as a {\em label} for these features, but the reader is requested
to forget the literal meaning, as it is probably quite misleading.  In
our usage, `ring' means the whole feature which in several cases
contains significant internal structure as well as a rim; in some
cases the rim-brightening is not very obvious.

We identify five of these ring features, which we have labelled E
through A, with their rough outer boundaries sketched in
Fig.~\ref{xjets}.  In most cases there is a bright `head' to the ring
on the side more distant from the core, which we label Ah, Ch, Dh, and
Eh. Eh is itself a `mini-ring', but we consider this part of ring E
because their outer boundaries are continuous.  Although several
authors have claimed that the ring features are genuinely circular or
elliptical, in fact ring C is the only one whose bright rim extends
(nearly) all the way round the structure. It is tempting to see ring B
as a circular feature with an additional component projecting out at
PA 45\degr, but close inspection shows no sign of a circular rim in
this quadrant, and B is best described as a rough oval centred
substantially off the path defined by the other rings. The western rim
of the final ring A is a prominent semi-circular arc (Ah). This is set
back slightly from the outer edge of A along the sections trailing
towards the east, giving an apparent double edge (see also
Fig.~\ref{xtoti}).

Our terminology
follows DF84; the reader should note that the five rings described by
\citet{Mason.etal1988} and \citet{Morrison.etal1996} differ from ours:
their first corresponds to Eh rather than the whole of E, they
do not identify D, their third and fourth rings are fitted to the 
smoothly-curved south-west and western segments of our B and A, ignoring
the rest of these features, and their fifth ring is the outer edge
of the lobe. 

On the east side the centre-brightened jet co-exists with
rim-brightened features F through J analogous to the rings on the west
side; these are labelled in Fig.~\ref{xjets}.  Whereas the counter-jet
appears to vanish at W4, the main jet brightens dramatically beyond E4
into the complex E5--E6 region, in which the jet curves to the south by about
12\degr. Beyond E6 the jet narrows and fades until the
highly-collimated part terminates at knot E8 which is elongated
perpendicular to the jet.  The faint ring J connects directly to E8.
Beyond E8 the jet flares in
width and no clear ridge-line can be followed through the complex
knots E9--E10.  The structure of E8 and ring J is repeated in E11 and
ring I.  The southern rim of I joins the jet
downstream of E11, so may not be part of a coherent `ring'.  Beyond
E11 the jet is surrounded by `ring H', which is complex and
may represent more than one ring-type feature. In this segment the jet
appears to re-collimate, but this is most likely an illusion
because somewhere in this region it bifurcates into two strands which
terminate in the blunt knots E12 and E13 respectively. 
Ring G, surrounding E13, seems to be
embedded in the final section of the jet, the broad fan of material F;
in particular, the edge of F clearly continues to the north of ring G
in Fig.~\ref{xjets}.

\subsubsection{The core}

The core is unresolved in all our images, with an upper limit of 0.07 arcsec
from Gaussian fitting to the full resolution 4848 MHz image.
Its best-fit position, from the better-calibrated 8440 MHz data, is  
16\hr 48\mn 39\fs 960, 05\degr 04\arcmin 35\farcs 318.
Fluxes at each frequency are given in Table~\ref{core}.

\begin{table}
\caption{Radio core}
\label{core} 
\begin{tabular}{lr} \hline
8440 MHz &  4.52 mJy \\
4848 MHz &  9.99 mJy \\
1665 MHz & 32.7~ mJy \\ 
1435 MHz & 39.2~ mJy \\
1365 MHz & 41.5~ mJy \\
1295 MHz & 43.0~ mJy \\ \hline
\end{tabular}
\end{table}

There is no sign of variability.  In the four months between the A and
B configuration observations at 4848~MHz, the core flux changed by $\la
0.1$ mJy. Also in the three months between the B and C configuration
observations at 8440~MHz the core varied by less than 3\%. 

From the fact that it is optically thin down to 1295 MHz, we infer that the
brightness temperature is less than $10^{12}$~K. This gives a lower limit
to the core diameter of 0.2 mas.  Observationally, maximum brightness 
temperatures are usually around $10^{11}$~K \citep{Readhead1994}, suggesting
a size $\ga 0.7$~mas.  In fact new EVN and MERLIN observations of the 
core region \citep[and in preparation]{GGL} revealed 
emission elongated in the NW-SE direction on 10--20 mas scales,
substantially misaligned with the kpc-scale jets.

\subsection{Spectral Index}
\label{14spix}

\begin{figure*}
\centering
\setlength{\unitlength}{1cm}

\begin{picture}(17.5,9.5)
\put(-0.3,10){\includegraphics{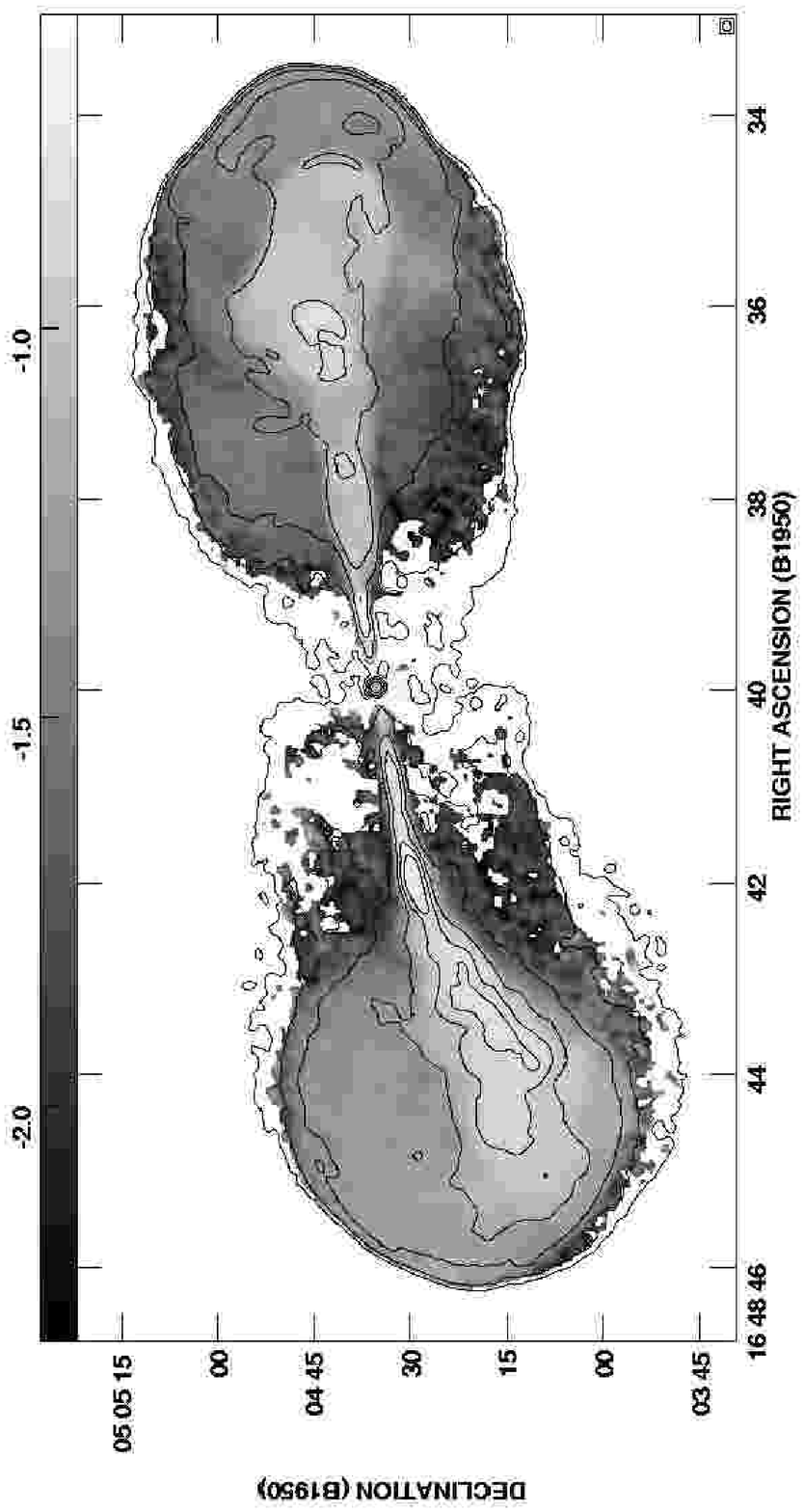}}
\end{picture}

\caption{Grey-scale of the spectral index map between 1.3 and 4.8 GHz
at 1.4 arcsec resolution. 
The grey-scale runs from $-2.3 \le \alpha \le -0.6$. The
flattest value is $\alpha = -0.61$. Only a few pixels (with large errors)
have $\alpha < -2.2$.
Contours are of the stacked 1.4 GHz map, separated by
factors of 3 starting at 0.5 mJy beam$^{-1}$. The mean error on displayed
points is 0.07, with a full range from .01 to 0.3.}
\label{spixm} 
\end{figure*}

Fig.~\ref{spixm} shows the image
of $\alpha^{4.8}_{1.3}$. The image at $\alpha^{8.4}_{4.8}$ is qualitatively
similar, except that the bridge is too faint at 8 GHz for its
spectral index to be measured.

The spectral index image dramatically highlights the distinction between
the bright and diffuse structures that we made in the previous section.
The former have much flatter spectra with a particularly sharp spectral
boundary in the western lobe. In the east the edge of the high-brightness
features is also a sharp edge in spectral index, except for the fan
component (F) at the end of the jet, in which the spectrum appears to
steepen smoothly into the lobe. There is also a rather abrupt separation
between bulb and bridge on the eastern side (most of the western
bridge is too faint to measure $\alpha$). In particular, patches with
$\alpha \approx -2$ around the southern
edge of the eastern lobe are part of the bridge, c.f. Fig.~\ref{20toti}.
For brevity we will speak of the flat, steep, and very steep spectrum 
components, as there is no truly flat spectrum 
structure in Her~A.

Table~\ref{spix} shows the average value of the spectral
index in various regions of the source. As can be seen
also from Fig.~\ref{spixm} the spectra of all parts of
Her~A are unusually steep.
Even more interesting is that the core is not only optically thin,
but has a steeper spectrum than the bright jet knots.

\begin{table}
\caption{Averaged values of the spectral index}
\begin{center}
\begin{tabular}{lll} \hline
Region & $\alpha(1.3,4.8)$ & $\alpha(4.8,8.4)$ \\ \hline
Total & $-1.42$  & $-1.56$ \\
East side & $-1.39$ & $-1.45$ \\
West side & $-1.46$ & $-1.65$ \\
East bulb$^a$ & $-1.31$ &  $-1.65$ \\
East Bridge & $-1.84$ & --- \\
East Jet & $-1.02$ &  $-1.13$   \\
West bulb$^a$ & $-1.51$ & $-1.83$ \\
West Bridge & $-1.92$ & --- \\
West Jet/Rings & $-1.09$ & $-1.25$ \\
Core  & $-1.12\pm0.02$ & $-1.41\pm0.04$ \\ \hline
\end{tabular}
\end{center}
{$^a$ \footnotesize Excluding the jets and rings.}
\label{spix}
\end{table}

The spectral index of the jets and rings is contaminated by the lobe
material superposed along the line of sight, and so the apparent
spectral index of the high-brightness features is steeper than the
true value. The spectral tomography gives a better idea of the true
values.  The results for various features are listed in
Table~\ref{tomo}; in some cases (e.g. ring E) there is significant
internal variation in $\alpha$, but for simplicity we just list
typical values.

\begin{table}
\caption{Spectral indices measured from spectral tomography}
\label{tomo}
\begin{tabular}{lcc} \hline
Component & $\alpha(1.3,4.8)$ & $\alpha(4.8,8.4)$ \\ \hline
W1--W2 & $-0.59 \pm 0.02$ & $-0.72 \pm 0.03$ \\
W4     & $-0.60 \pm 0.05$ & $-0.76 \pm 0.04$ \\
E      & $-0.85 \pm 0.03$ & $-1.04 \pm 0.05$ \\
Eh     & $-0.85 \pm 0.03$ & $-0.98 \pm 0.03$ \\
D      & $-0.75 \pm 0.05$ & $-0.98 \pm 0.03$ \\
Dh     & $-0.75 \pm 0.03$ & $-0.70 \pm 0.03$ \\
C      & $-0.77 \pm 0.03$ & $-1.06 \pm 0.04$ \\
Ch     & $-0.74 \pm 0.02$ & $-0.94 \pm 0.03$ \\
B      & $-0.82 \pm 0.03$ & $-1.04 \pm 0.03$ \\
A      & $-0.75 \pm 0.05$ & $-1.04 \pm 0.03$ \\
Ah (Arc) & $-0.90 \pm 0.05$ & $-1.14 \pm 0.04$ \\
E1--E3 & $-0.76 \pm 0.02$ & $-0.90 \pm 0.03$ \\
E4     & $-0.61 \pm 0.02$ & $-0.76 \pm 0.03$ \\
E5--E6 & $-0.75 \pm 0.01$ & $-0.87 \pm 0.02$ \\
E7--E8 & $-0.80 \pm 0.03$ & $-0.90 \pm 0.03$ \\
E9     & $-0.79 \pm 0.02$ & $-0.87 \pm 0.03$ \\
E10--E11 & $-0.77 \pm 0.02$ & $-0.88 \pm 0.04$ \\
E12--E13 & $-0.80 \pm 0.03$ & $-0.92 \pm 0.03$ \\
F--H  & $-0.85 \pm 0.03$ & $-1.02 \pm 0.04$ \\
I     & $-0.76 \pm 0.02$ & $-1.04 \pm 0.06$ \\ \hline
\end{tabular}
\end{table}

The tomography results show that the flattest-spectrum regions, with
$\alpha^{4.8}_{1.3} \approx -0.6$ are the western jet and knot E4 in
the eastern jet. The rest of the eastern jet has 
$\alpha^{4.8}_{1.3} \approx -0.75$, steepening to $-0.8$ beyond E8, 
only a little steeper than typical for bright jets ($\approx -0.6$), and
close to typical values for hotspots, $\approx -0.75$ 
\citep{Alexander.etal1987}.
Rings A, C, D and I have similar spectral indices to the eastern jets,
while rings B, E, F, and H are a little steeper at 
$\alpha^{4.8}_{1.3} \approx -0.85$. 
The bright western
arc Ah has the steepest spectrum of any feature in the jet/ring system. 
It is situated
on the boundary between the flat spectrum material in feature A
and a steep-spectrum filament (see below), which makes 
the spectral tomography difficult to interpret, but it is
clear that the spectrum of Ah is inconsistent with either of these components.

\begin{figure}
\centering
\setlength{\unitlength}{1cm}
\begin{picture}(8.5,5.0)
\put(-0.2,5.2){\includegraphics{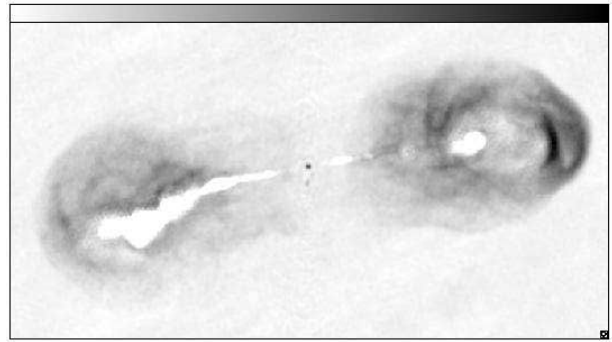}}
\end{picture}
\caption{Tomography image (see Section~\ref{alpha_analysis}) for
$\alpha^{4.8}_{1.3}= -0.9$, i.e. any feature with a flatter spectrum
(including essentially all of the jet and ring components) appears
negative (light).  The positive feature south of the core is the
phase-centre artefact in the 5-GHz image.}
\label{tomomap}
\end{figure}

Fig.~\ref{tomomap} shows the spectral tomography image for 
$\alpha^{4.8}_{1.3}= -0.9$; that is, with the maps scaled so that emission
with this spectral index is exactly subtracted out. 
Here essentially all the jet and ring components
have been over-subtracted.  Features in this image may represent a combination
of intensity and spectral index structure. Comparison with Fig.~\ref{xtoti}
shows that most of the filamentary features in the lobes visible in
Fig.~\ref{tomomap} are intensity
features, with the exception of the brightest. This is {\em not}
the arc Ah, which is almost exactly subtracted here, but a filament of
steep-spectrum emission just beyond it. With $\alpha \approx -1.5$, its
spectrum is significantly steeper
than most of the emission in the western bulb.
The emission remaining in Fig.~\ref{tomomap} seems to be
organised around the flat spectrum material. In the
western lobe this takes the form of filaments arching around the ABC complex.
In the eastern lobe the organisation is less coherent, but the brightest
parts of the steep-spectrum emission are all close to the jet.

\begin{figure}
\centering
\setlength{\unitlength}{1cm}
\begin{picture}(8.5,8.0)
\put(-0.5,-0.5){\includegraphics{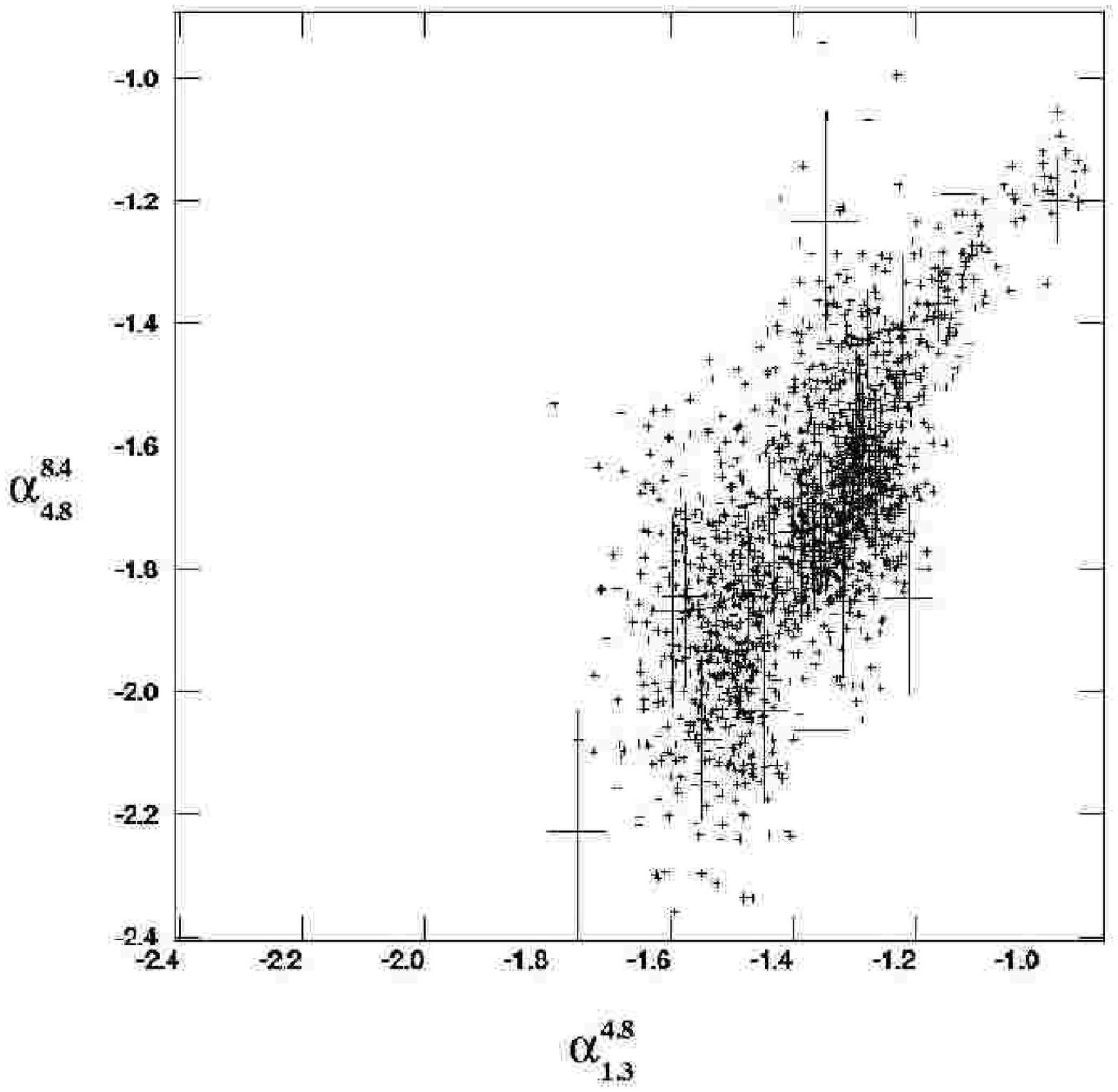}}
\end{picture}
\caption{Radio colour-colour plot of the lobes of Her~A (the jets
and rings have been masked out of the images used to make this plot). 
The abscissa is $\alpha^{4.8}_{1.3}$; the ordinate is $\alpha^{8.4}_{4.8}$.
Colours are plotted for image pixels separated by 1.2 arcsec, 
so they are nearly independent. 
Points are only plotted if the error is less than 0.2 in both $\alpha$ values. 
Error bars are plotted on every 50th point; note that unusually large errors 
are the most apparent.}
\label{colcol}
\end{figure}

As expected for aged synchrotron emission, the spectra of all components
(except the faint knot Dh) steepen at high frequency. 
The steepening is mild in the jets and the compact heads of the western rings, 
typically $\Delta\alpha \approx 0.12$. All the rings show more steepening
than the jets, typically $\Delta\alpha \approx 0.23$. Steepening is
clearest in the bulbs, where the curvature can even be detected in the 
$\chi^2$ for the 1.3--4.8-GHz fits. 
Fig.~\ref{colcol} shows a radio `colour-colour' plot, with points from
the jet and core region excluded. Because the very steep spectrum
bridge is mostly not detected at 8 GHz, it is also 
effectively excluded.  The distribution of
points resembles that found for Cygnus A by \citet*{Katz-Stone.etal1993};
while curvature increases with spectral index, as expected from the
simple \citet{Jaffe.etal1973} spectral ageing model, it does not increase
as fast as predicted.

\subsection{Fractional Polarization}
\label{fpol}

\begin{figure*}
 
\centering
\setlength{\unitlength}{1cm}  
 
\begin{picture}(17.5,8)
\put(-0.3,8.5){\includegraphics{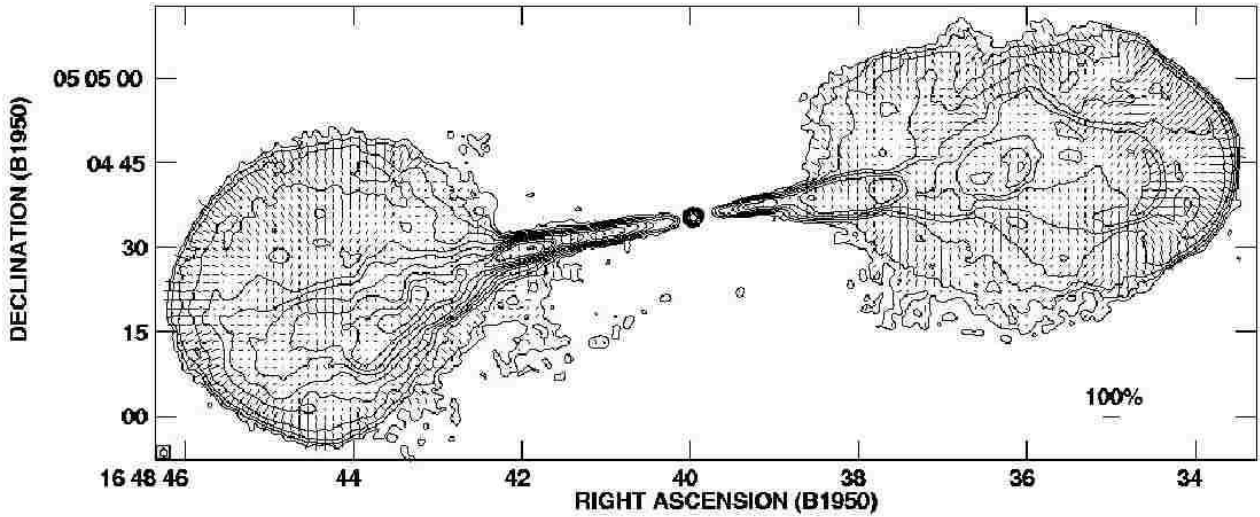}}
\end{picture}

\caption{The fractional polarization map of Her~A at 8440 MHz at
1.4 arcsec resolution.  Line segment orientation gives the E-vector
direction and the length is proportional to the fractional polarization.
Contours are from the total intensity map and are
separated by a factor of two. The bar labelled `100\%' gives the scale
for the vectors, which are
only plotted within the outer contour (0.07 mJy beam$^{-1}$), 
and when the signal-to-noise is greater than 2.}
\label{3.5fpn}
\end{figure*}

\begin{figure*}
 
\centering
\setlength{\unitlength}{1cm}

\begin{picture}(17.5,8)
\put(-0.25,9){\includegraphics{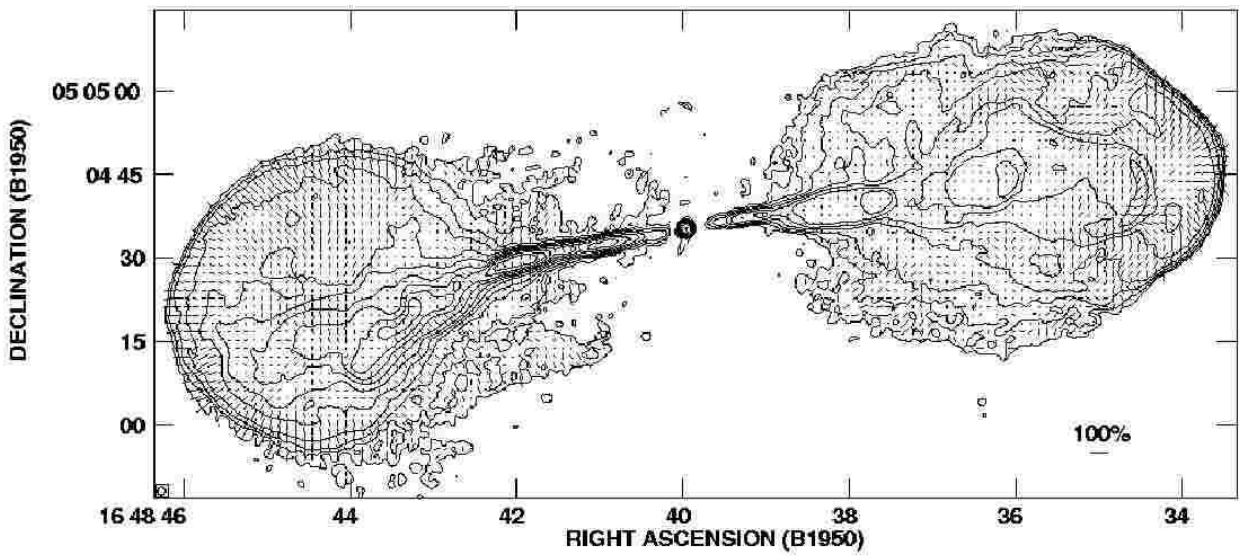}}
\end{picture}
\caption{Fractional polarization map at 4848 MHz. 
Details as for Fig.~\ref{3.5fpn}, except that the outer contour is at
0.14 mJy beam$^{-1}$.}

\label{fp1}
\end{figure*}

\begin{figure*}
 
\centering
\setlength{\unitlength}{1cm}

\begin{picture}(17.5,8)
\put(-0.3,8.8){\includegraphics{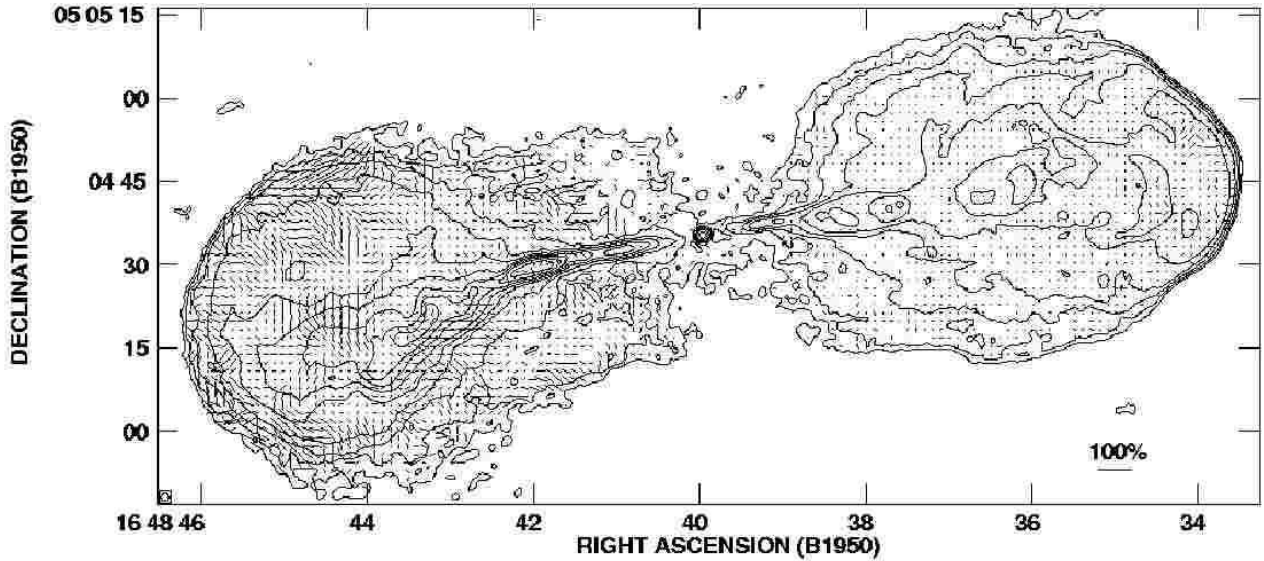}}
\end{picture}
\caption{Fractional polarization map at 1665 MHz. Details as in
Fig.~\ref{3.5fpn} except that the outer contour is at 0.66 mJy
beam$^{-1}$. Note that the vector scale is double that in the previous
plots.}
\label{18fp}
\end{figure*}

The fractional polarization maps at 8440, 4848 and 1665~MHz 
are presented in Figs~\ref{3.5fpn}, \ref{fp1} and \ref{18fp} 
respectively. These maps reveal strong asymmetrical Faraday
depolarization with the eastern side, containing the stronger jet, being
less affected. Therefore Her~A exhibits a strong Laing-Garrington
effect \citep{Laing1988,Garrington.etal1988}.

On the western side, at 8 GHz the E-vectors are mainly orthogonal to
the edges of the ring-like features and the edges of the lobes, but at
4.8 GHz they begin to become disordered especially near the centre,
and the degree of polarization begins to drop. At 1.7 GHz and lower
frequencies, the fractional polarization is barely detectable over
most of the lobe, except that some weak polarization remains at the
extreme western end of the lobe.
 
On the eastern side, the vectors in the bulb remain ordered around the
lobe edges, and the edges of the eastern jet, down to 4.8 GHz. In the
faint eastern bridge emission, the 4.8-GHz polarization is noisy, but
is at least roughly orthogonal to the edges of the lobe. At 1.7 MHz
and below, the lobe polarization becomes disordered and generally
depolarizes, although at some isolated points the degree of
polarization at least temporarily increases.

\begin{figure*}
\centering
\setlength{\unitlength}{1cm}
\begin{picture}(17.5,5.5) 
\put(-0.3,6){\includegraphics{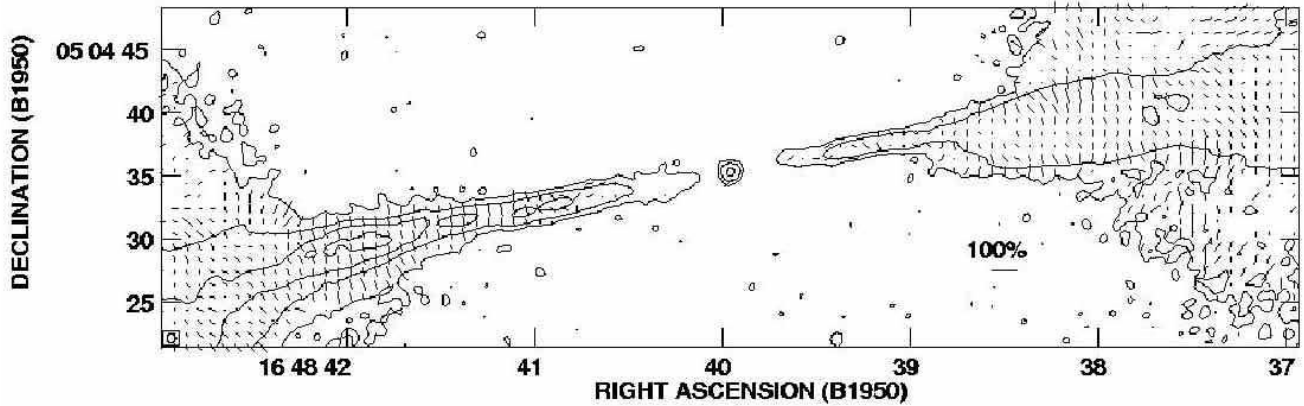}}
\end{picture}
\caption{Fractional polarization map of the inner jets at 8440~MHz at
0.74 arcsec resolution. Contours are separated by factors of eight.}
\label{high_xfp}
\end{figure*}

Fig.~\ref{high_xfp} presents the distribution of the fractional
polarization vectors in the inner jets at 8440~MHz at the higher
resolution of 0.74 arcsec.

\begin{figure}

\centering
\setlength{\unitlength}{1cm}

\begin{picture}(10,9)

\put(-2.3,-7.5){\includegraphics{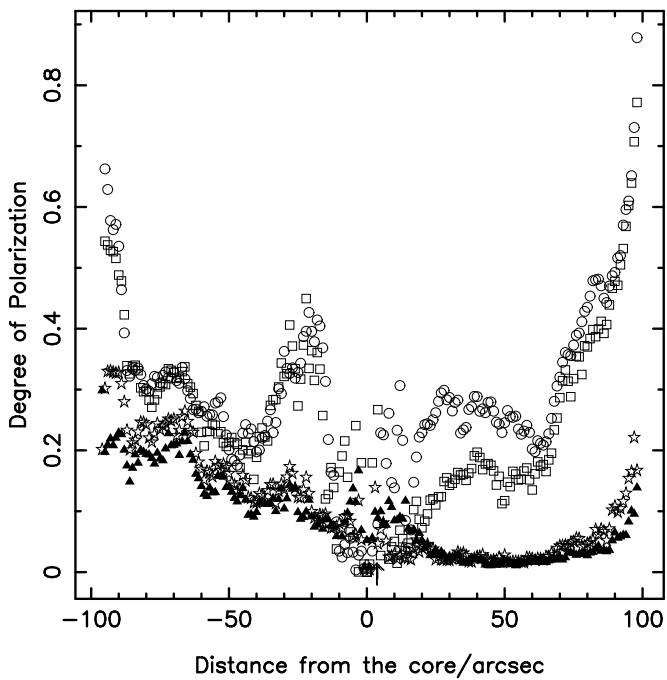}}
\end{picture} 
\caption{Strip averages of degree of polarization $m$ plotted against
distance from the core at 1.4 arcsec resolution.  Filled triangles:
1295 MHz, stars: 1665 MHz, squares: 4848 MHz, circles: 8440
MHz. Negative distances represent the eastern side. The `arrow'
symbol denotes the position of the optical companion located 4 arcsec
($\simeq$ 11 kpc) north-west from the nucleus of Her~A and could have
any depolarization value.}
\label{miring}
\end{figure}

To quantify the dramatic changes in the polarization from one end of
the source to the other, Fig.~\ref{miring} plots `strip averages' of
the polarization across the source against position along the source
axis. More precisely, we divided each lobe into concentric ring
segments centred on the radio core, of width 1 arcsec and radius up to
120 arcsec.  In each ring we averaged the fractional polarization $m$
at 1295, 1665, 4848 and 8440 MHz from the maps. In the plots, negative
distances denote the eastern side of the source (lobe and jet) and
positive ones the western side.  Note that near the centre the 1295
and 1665 MHz values reflect the faint inner bridge, but this is not
detected at 4848 and 8440 MHz so the inner jets dominate at these
wavelengths. Thus the curves cannot be directly compared in the inner
$\approx 20$ arcsec.

This plot emphasises the dramatic Laing-Garrington effect in the source,
which in Paper III we interpret as primarily due to external depolarization
by the magnetic field in the cluster halo, with the western lobe on the
far side of the cluster core and the eastern lobe in front.
Some depolarization should be
caused by the optical companion (its position is denoted by the
`arrow' symbol in Fig.~\ref{miring}) and by nearby galaxies. These
points will be discussed in Paper III, where we present our rotation
measure and depolarization maps.

\subsection{Projected Magnetic Field}

\begin{figure*}
\centering
\setlength{\unitlength}{1cm}
  
\begin{picture}(17.5,22)
\put(3,-0.5){\includegraphics{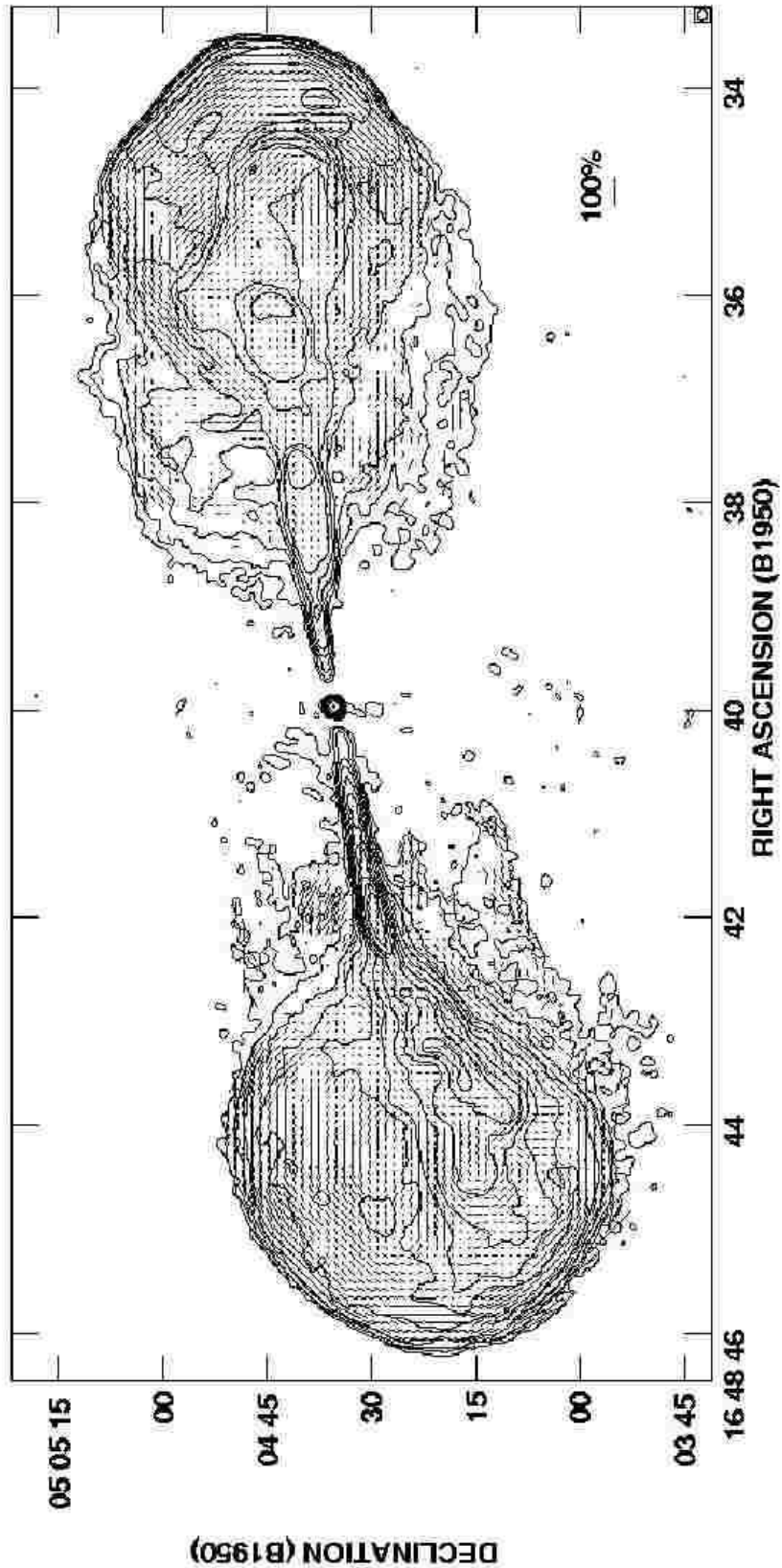}}
\end{picture}

\caption{Projected $B$-field directions at 1.4 arcsec
resolution. Vector length is proportional to $m(5)$ and the contours
show total intensity at 5 GHz.  Contours are plotted above 0.145 mJy
beam$^{-1}$ (5$\sigma$), separated by a factor of two. 
The $B$-vectors are at
90\degr\ (roughly) to the fractional polarization vectors.}
\label{bvm}

\end{figure*}

Our estimate of the $B$-field map is shown in Fig.~\ref{bvm}.
The faint inner bridge is only
detected at 1.4 GHz, where Faraday rotation and depolarization are 
large; so that even when polarized flux is detected (usually at 1665 MHz), 
the Faraday correction to zero wavelength is very uncertain.  
For most of these pixels we therefore cannot determine a reliable $B$-field.  
As noted in Section~\ref{pol_analysis} 
we had to resolve ambiguities in rotation by
hand in the western lobe. In practice
we have taken $n\pi = 0$ over the outer $\sim$ third of the lobe, which
seems justified given the clear alignment of the uncorrected polarization 
pattern with the lobe boundary at both 8 and 5 GHz  
(Figs~\ref{3.5fpn} \& \ref{fp1}).

The projected magnetic
field closely follows the edge of the source, the jets, and the
ring-like structures in the lobes; the field pattern in the two lobes
is broadly similar.  The overall trend for alignment of the magnetic field
with edges in the emission is very typical for DRAGNs \citep*[e.g.][]{LPR86}
although the high quality of our data reveals this particularly well.

\section{Discussion}
\label{discussion}

\subsection{Spectral index asymmetry}
\label{alpha_prof}

The fact that in powerful DRAGNs the less depolarized lobe has a
flatter spectrum has been observed by \citet{Liu.etal1991},
\citet{Garrington.etal1991a} and \citet{Garrington.etal1991}.  There
has been a debate concerning the origin of depolarization and spectral
asymmetries between the two lobes of DRAGNs, especially because
\citet{Pedelty.etal1989} found that depolarization was systematically
stronger on the shorter side, which suggested an intrinsic effect
rather than an orientation one as advocated by
\citet{Garrington.etal1991}.  Explanations have been given either in
terms of asymmetric environment \citep{Fraix1992}, or in terms of an
intrinsic asymmetry in the DRAGN, perhaps in the kinetic power of the two
jets, and/or the strength of the $B$-field in the lobes
\citep{Alexander1993}, or by projection
effects alone, provided that the lobe magnetic flux decreases as
the DRAGN evolves and that the DRAGN axis is no more than 20\degr\
from the plane of the sky \citep{Blundell.etal1994}.

A detailed study of ten quasars by \citet{Den.etal1997}
found that the high-brightness material obeyed the Liu-Pooley relation
while fainter emission followed the Pedelty et al. result. In their
interpretation, Doppler beaming boosted bright flat-spectrum emission
in the approaching lobe, but at fainter levels intrinsic asymmetries
dominate.  Her~A provides an interesting test of this model.

\begin{figure}

\centering
\setlength{\unitlength}{1cm}
 
\begin{picture}(10,8.5)
 
\put(-0.9,-0.5){\includegraphics{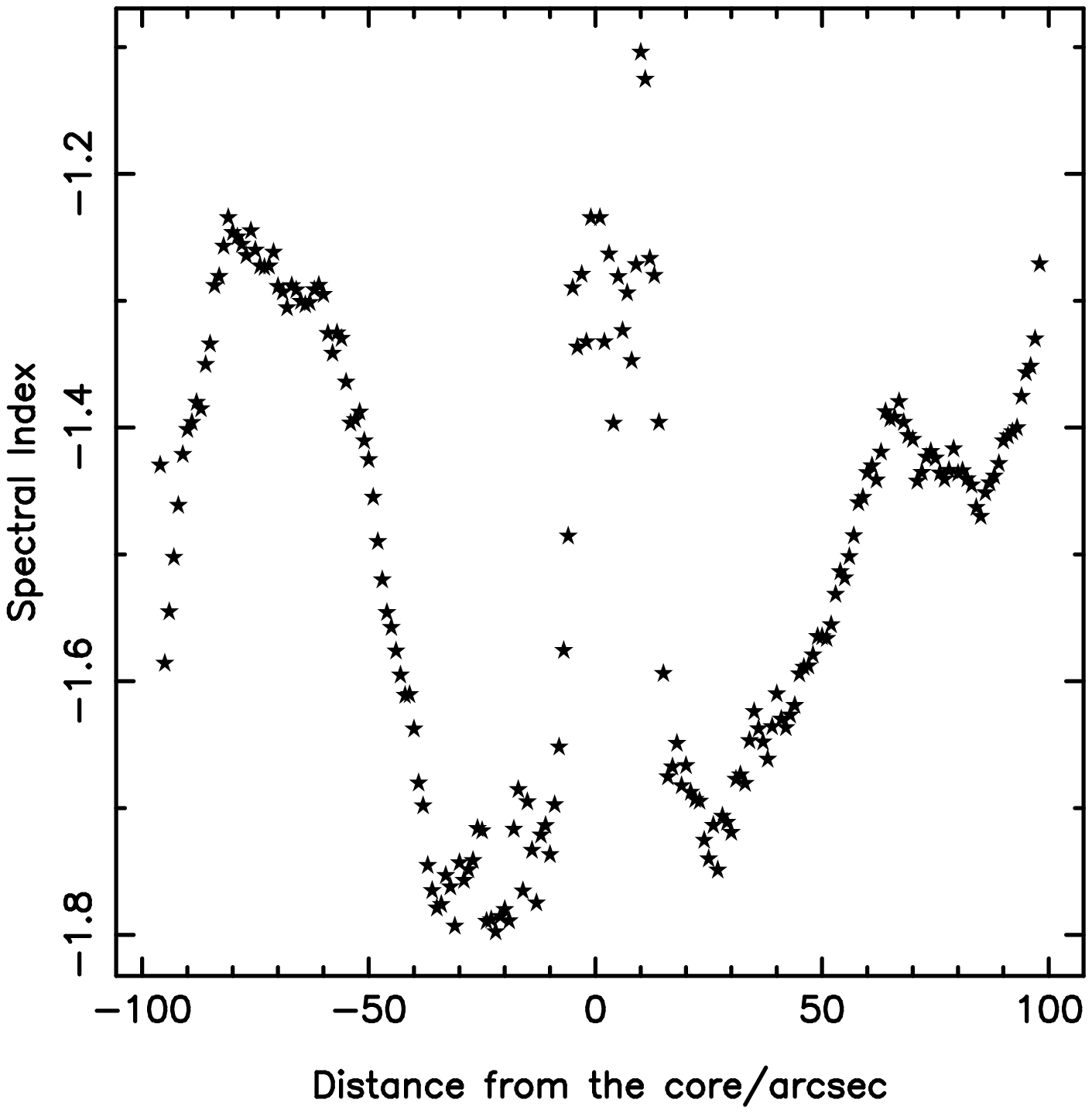}}
\end{picture} 
\caption{Plot of strip averages of the spectral index $\alpha$ at 1.4
arcsec resolution against distance from the core.}
\label{spring} 
\end{figure}

Fig.~\ref{spring} shows strip averages of the spectral index, obtained
in the same way described earlier. This is less useful than for
depolarization because much of the structure in $\alpha$ is parallel
to the DRAGN axis, so one can get a clearer and more detailed idea
from the spectral index map (Fig~\ref{spixm}).

From Figs~\ref{spring} (cf. also Fig.~\ref{spixm}) we can see that:

\begin{itemize}
\item At the ends of the source the eastern side has a steeper spectrum
than the western one. Steep
indices located towards the outer borders have been observed also in
Centaurus A and in FR\,Is in general \citep[e.g.][]{Compi.etal1997}.

\item Moving from the eastern ends towards the core, the lobe (and
partly the jet) spectra become flatter (from $\approx 90$ to $60$
arcsec east from the core) with $\alpha \approx -1.25$. The western
lobe, which is more depolarized, shows a steeper spectrum than the
eastern lobe. 
We derived the same result by comparing the depolarization
maps (see Paper III) with the spectral index map (Fig.~\ref{spixm}),
and also the plots of the degree of polarization versus distance
(Fig.~\ref{miring}) with the one for the spectral index
(Fig.~\ref{spring}).

\item At $\simeq 40$ arcsec east from the core the spectral index
starts steepening towards the core. This is because the steep spectrum
areas of extended emission, which contribute to the `strip averages',
are becoming apparent at 1.4 arcsec resolution (see Figs~\ref{spixm}
and \ref{20toti}). This trend reverses close to the core, especially
on the western side, because the steep-spectrum material becomes too
faint for us to determine spectral indices, leaving the flat spectrum
jets to dominate again.

\item Moving away from the core and into the western side the spectrum
becomes steep again (due to the steep spectrum extended emission);
$\approx 25$ arcsec  west from the core $\alpha \approx -1.7$.

\item There is a local maximum of the spectral index at 60 arcsec west
from the core. This is because between 25 and 80 arcsec the 'strip
average' have contributions from both the inner lobes (rings) and the
old lobes. The rings get wider and wider until $\sim$ 60 arcsec (rings
C \& B, see Fig.~\ref{xjets}) and this makes the average spectral
index flatter. From 60 to 80 arcsec the rings narrow to nothing and so
the old lobes become more and more important, cancelling out the fact
that in the old lobes themselves, the spectrum steadily flattens as
you move away from the core. At $\simeq 80$ arcsec west from the core
the averaged spectrum steepens because the flat spectrum region around
the rings comes to an end, giving a maximum $\alpha\simeq
-1.45$. Finally the most distant edges of the source appear flatter
than the ends of the eastern side as mentioned before.
\end{itemize}

The sample which Liu \& Pooley (1991) studied consisted of powerful
radio sources, most of which show no jet, or a jet with flux a
negligible fraction of the whole.  For this reason they argued that
the line-of-sight effects which can explain the Laing-Garrington
effect cannot account for the depolarization-spectral index
correlation. In the case of Her~A (see Table~\ref{spix}) the flatter
spectrum eastern jet dominates the flux of the surrounding
steeper-spectrum lobe, so it is no surprise that the Liu-Pooley
correlation holds when the spectral index is assessed for the entire
lobe, including the jet. But the spectral index asymmetry is clearly
present in the `old' lobe material, and so, at least in this case,
cannot be explained as an orientation effect. Our discussion in
Section~\ref{restarting} explains the injection of new,
flatter-spectrum material into the jet-side lobe in terms of the time
delay across the source, but this explanation cannot work for most
DRAGNs; only a small fraction of them will be observed shortly after
the jets re-start. We note that Blundell \& Alexander (1994) give a
related explanation which {\em can} work in most cases.

In contrast, models which explain the asymmetry in terms of intrinsic 
differences between
the lobes, reflecting either long-term asymmetries between the two jets, 
or some large-scale asymmetry in the environment, face a
problem in Her~A because of the almost-exactly equal size of the two lobes.
Moreover, our X-ray images (Paper I) do not suggest a strongly asymmetric
environment.

\subsection{The bridges}
\label{bridge}

The bridge emission in the centre of Her~A poses a puzzle. These are
the faintest parts of the source, which at first sight suggests the
lowest emissivity and so the lowest internal pressure, and yet the
bridges seem to occupy the cluster centre, where the thermal pressure
in the environment is the highest. Similar faint bridges are found in
many powerful DRAGNs \citep*{Leahy5.etal1989}, but only in Her~A and
Cyg A can we compare the radio and X-ray structures (see Paper I).  In
Her~A, the bridges are (in projection, at any rate) within the core of the
thermal gas distribution of the cluster. Unlike the case of Cyg A,
there is no hint of X-ray `holes' indicating that the intra-cluster
medium (ICM) has been displaced by the bridges. Thus if the source is
roughly cylindrical, the radio-emitting region must be well mixed with
the ICM.  But in that case we would expect that after such mixing the
concentration of relativistic plasma would decline smoothly with
distance from the source axis (marked by the jets) whereas in fact the
edges of the bridges are sharply-defined in Fig.~\ref{20toti},
especially the eastern bridge which is actually edge-brightened. In
addition, mixing should give turbulent eddies along the surface of the
bridges and/or bulb edges in Her~A, but there is no sign of such
vortices.

In our spectral index map the bridges are the steepest-spectrum regions
in Her~A, and so at lower frequencies the brightness contrast between
bridge and bulb will be less marked.  One might be tempted to assume
that, in an image at the lowest frequencies, the distinction between 
bulb
and bridge will vanish altogether. This would allow roughly constant internal
pressure, since the pressure is carried mainly by the electrons which 
radiate at the lowest frequencies. However, 
\citet{Kassim.etal1993} show that the bridge region 
remains faint even at 74 MHz, and this is typical of other DRAGNs.
As in other cases, the spectra cannot be extrapolated to low frequencies 
as power laws, because most of the spectral index structure reflects
curvature of the spectrum in the GHz band.  

A plausible model is that the bridges are only projected onto the
cluster centre (and to some extent, onto each other).  This makes it
easier to understand how the clear division between the two bridges
can be maintained.  A narrow gap between the bridges is seen in some
DRAGNs \citep{LP91,Johnson.etal1995}, which from many viewing angles
would be closed in projection. Thus in Her~A, the bridge material 
consists of a skirt of
emission displaced in front of and behind the core, 
The thickness of the bridges along the line of sight could then be
smaller than their width, 
so that their emissivity (and  pressure) would be closer to those of the 
bulbs than the images initially suggest. This model also makes some sense
of the edge-brightening of the eastern bridge.

On this interpretation of Her~A, the
inner jets are confined by the thermal gas in the cluster core,\footnote{%
As described in Section~\ref{jetwidth}, there is strong evidence that
the western jet is temporarily confined by ram pressure due to lateral
expansion; but this should be a relatively brief episode and our model
here applies to the more normal state of affairs.}
and it seems likely that the transitions in the jets near knots E4 and W4
mark the points of entry into the bulbs, as suggested by
\citet{Meier.etal1991}.  Our X-ray analysis (Paper I) gives a thermal 
pressure in the cluster core of 13~pPa, while the equipartition pressure
in the eastern jet is $\sim$2.3 pPa with the standard assumptions (including
no protons).  These values are certainly compatible with thermal confinement
\citep[c.f.][]{Leahy.etal1999}.

\subsection{Collimation}
\label{jetwidth}

Fig.~\ref{collim} shows that the inner jets have approximately constant
width when they are bright enough to measure easily. The innermost parts 
of the jets (within $\approx 8$--10 arcsec of the core) are very faint,
so width measurements become uncertain, but there is weak evidence that 
the jets narrow towards the core as expected.
The broadening of the eastern jet at $-6$ arcsec may be because
the fits are affected by structure in the surrounding bridge. 
The initial FWHM opening angle ($\arctan d \theta_J/d l$)
must be greater than 3\degr--4\degr.
Rapid opening followed by near-constant width is is typical of 
strong-flavour jets \citep*{BPH1986,LP1995,SBB1998}.

The western jet is obviously much narrower than the eastern.  It is
best to compare the western jet measured at 5 GHz with the eastern at
8 GHz, because the ratio of jet width to beam width is then similar;
we find 0.40 ($\simeq$ 1 kpc) and 0.78 arcsec ($\simeq$ 2 kpc) for the
west and east widths.

Comparing the flux per unit length in the E1--E2 region with the
W1--W2 region (see Fig.~\ref{ABCDjets}) the ratio is 2.6:1; thus the
emissivity (uncorrected for beaming) in the western jet is {\em
higher} than in the eastern by a factor of 1.5. 
understand the difference in emissivity in terms of adiabatic
expansion. For highly supersonic jets the speed is nearly constant
along the jet, and so only transverse expansion is important.  If the
jet expands by a factor $\cal R_\perp$ in each direction transverse to
the jet flow, the emissivity should fall as ${\cal R}_\perp^{-(12 +
10\alpha)/3}$ if the magnetic field is parallel to the flow, and as
${\cal R}_\perp^{-(10 + 8\alpha)/3}$ if the field is tangled
\citep{Leahy1991}.  In fact the eastern jet shows 20--45 per cent
polarization at high frequency, with the field along the jet
(Fig.~\ref{high_xfp}), suggesting a reasonably organized but not
purely parallel field, so we expect behaviour somewhere between these
two extremes.  If we assume the two jets are initially identical, they
differ in expansion at $\sim$ 12 arcsec from the nucleus by ${\cal
R}_\perp = 1.95$, and with $\alpha$ between $-0.6$ and $-0.75$ we find
the narrower jet should have an emissivity higher by a factor between
27 and 77.  It is therefore quite possible that the western jet could
be Doppler dimmed by a factor of $\sim 20$ with respect to the
eastern, which is consistent with the values derived below for the
outer jets.

This leaves unexplained the difference in width between the two jets.
If our Doppler beaming estimate is correct, the western jet has an
intrinsic emissivity $\sim 30$ times higher, and hence its
equipartition pressure is $\approx 7$ times that in the
eastern. Adiabatic expansion arguments give a similarly large pressure
ratio (depending again on the assumed magnetic geometry).  Since the
jets have roughly constant width in this region, they must be confined
and so a similar pressure difference must exist in the confining
medium. It is not easy to see how this could be maintained statically
within the cluster core, and our X-ray images in Paper I do not show a
pronounced asymmetry (although a temperature difference might not be
very obvious). In Section~\ref{restarting} we discuss a dynamic model.

The eastern jet starts to flare at 23 arcsec, just after E3 (see
Fig.~\ref{ABCDjets}), with opening angle $\arctan(d\theta_J/dl) \approx 7\degr$
(the narrow
measured width at 26--28 arcsec reflects knots E4 and E5 rather than
the underlying jet). It reaches a maximum width in the bright section
between E5 and E6 before narrowing towards E8. Some way downstream of
the disruption point at E8, the effective opening angle is around
6\degr.  \citet{Meier.etal1991} interpret the flaring of the jet as
evidence that it becomes overpressured at around E3. In its simplest
form this concept is not consistent with the narrowing between E6 and
E8, but the jet is surely not in a steady state and so it is possible
that the overpressure is recent and did not affect the part of the jet
that has now reached E7--E8.

Could the broadening beyond E8 be due to the overpressure created by 
strong shocks at E8 and E11? The structure of these knots suggests 
shocks moving away from the core through the jet (c.f. Section~\ref{rings}), 
in which case material downstream
would not yet have been affected.  In this case these may be symptoms
of the disruption downstream, which would create an obstacle to the incoming
collimated flow.  But this interpretation may be over-simplified; the shocks
may be moving downstream more slowly than the underlying flow, allowing
them to influence the downstream jet. 

\subsection{Beaming and Symmetry}
\label{beaming}

The flux ratio between the flat-spectrum features in the eastern and
western lobe is $\approx 1.9$ at 1440 MHz, after correction for the
underlying steep-spectrum emission. At first sight this suggests that the
jets are not very asymmetric in flux, but very different features dominate
this ratio: in the west the rings, in the east the narrow jet itself.  
 
We cannot be absolutely sure that the western lobe contains a true jet
beyond W4, but there are several features which suggest that one does
exist, directly analogous to the eastern jet, but much fainter.
Figs~\ref{xtoti} and~\ref{xjets} (bottom) show that there are
relatively compact features (which have flat spectra) {\em within}
rings D and A, i.e. features which do not form part of the
edge-brightened rims. At 8 GHz, these are at the level of $\sim
0.2$--0.5 mJy beam$^{-1}$, ${\cal O}(10)$ times fainter than the jet
emission at corresponding distances in the eastern lobe. If we assume
that these features belong to a western jet beyond W4, similar in its
co-moving frame to the eastern jet, then, for $\alpha = -0.75$, this
difference could be attributed to Doppler beaming with $\beta \cos i
\approx 0.5$.  In Paper III we find from the depolarization asymmetry
that the inclination to the line of sight is $i \approx 50\degr$. This
value would give $\beta \approx 0.8$, consistent with the values
inferred for FR\,II jets from beaming arguments
\citep{Wardle.etal1997}.  We will use this combination of parameters
as a fiducial `fast model'.  With this value of $\beta \cos i$, a
feature in the western jet corresponding to the brightest part of the
eastern jet, at E5--E6, would have brightness $\sim 1$ mJy
beam$^{-1}$.  In fact at the same distance in the western lobe, the
elliptical ring contains fairly compact flat spectrum features at
about 0.6 mJy beam$^{-1}$, after correcting for underlying emission.
Thus, while the difference between the Her~A jets cannot be {\em
entirely} due to Doppler beaming, the intrinsic differences could be
much smaller than the images suggest, no greater than the observed
irregular brightness variations along the eastern jet.  This analysis
implicitly assumes that the jets are in a steady state, so that the
light-travel delay across the source does not prevent us from
comparing like with like. In contrast, we argue below that overall the
jet is highly variable. We excuse this contradiction on two grounds:
first, in at least the case of knots E5-E6 vs. the features in ring E,
it seems plausible that we are looking at structures associated with
the entry of the jets into the bulbs, i.e. at quasi-stationary
patterns, rather than moving `blobs' in the jet. Secondly,
relativistic beaming is a much more powerful source of asymmetry than
fluid-dynamical fluctuations in the jet flow, so that a systematic,
order-of-magnitude difference in jet brightness is likely to be {\em
predominantly} due to beaming.  Nevertheless, we will also consider a
`slow model' in which beaming is assumed to be negligible.  A problem
with a relativistic jet scenario 
is that the difference in width between the inner
jets cannot reasonably be produced by beaming, but this is addressed
in the follow section.  We have seen that the the emissivity ratio of
the inner jets, although at first sight in the wrong sense, is
consistent with beaming combined with differential adiabatic
expansion.

Previous authors starting with DF84 have claimed that the two lobes of
Her~A show several symmetries, including point-symmetric (S-type)
distortion of the jets/rings and paired structures at similar
distances on the two sides, from which they infer that nuclear
outbursts rather than instabilities are responsible for the major ring
and jet structures.  On close inspection the situation is more
complex. We have already noted (Section~\ref{bulbs}) that there are
systematic differences between the fine-scale structure of the two
bulbs (see also Fig.~\ref{tomomap}). As for the jets,
Fig.~\ref{ABCDjets} shows a very low amplitude S-distortion in the
inner jets, with the knots in the eastern jet being brightest on the
outside edges of the curves, as usual in jets.  On the other hand,
Fig.~\ref{xtoti} shows a large-scale mirror (C-type) symmetry. Around
this overall pattern, one can impose an S-type distortion by
associating ring B with jet knot E12, both of which project out of the
main envelope of the jet.  Despite the wandering of the outer boundary
of the rings, it is worth noting that knot W4 and the compact peaks Eh
and Ch are almost perfectly aligned with the core.

As for pairing features on the two sides, beyond the striking
coincidence of the bright region E5--E6 with the elliptical ring E on the
western side, there are rather too many features available in the eastern
jet to be sure of making the correct match. Table~\ref{correspondence} gives 
distances from the core for the major features we have identified,
showing a possible set of correspondences. 
By counting rings out from the
centre, we find that E12 and ring B do not seem to correspond to each other,
contrary to the association needed for S-symmetry.

\begin{table}
\caption{Possible correspondence between features in the two lobes}
\label{correspondence}
\begin{tabular}{llllr} \hline
East         & distance & West   & distance & ratio \\
feature      & arcsec   & feature & arcsec   & \\ \hline
E1           & 14.0  & W1 & 10.2 & 1.38 \\
E2           & 16.4  & W2 & 13.7 & 1.20 \\
E3           & 22.2  & W3 & 17.0 & 1.30 \\
E4           & 26.2  & W4 & 21.2 & 1.23 \\
J (E8)       & 42.5  & Eh & 36.0 & 1.18 \\
I (E11)      & 50.9  & Dh & 48.9 & 1.04 \\
H (E12)      & 64    & Ch & 59.0 & 1.08 \\
G            & 74.4  & B rim & 67 &  1.11 \\
F            & 85    & Ah & 82.1 &  1.04 \\ \hline
\end{tabular}
\end{table}

The light travel time from the back to the front of the source means
that we are seeing features in the nearer lobe at a later time in
their development. From the depolarization data
(Section~\ref{fpol}) we know that the eastern side is approaching,
so we would expect the ratio of east-to-west distances for
corresponding components to be larger than one, as in the scheme
proposed in Table~\ref{correspondence}.  The distance (or separation)
ratio should be given by $(1 + \beta \cos i)/(1 - \beta \cos i)$, so
the observed ratios suggest that $\beta \cos i$ falls from about 0.16
near the core to $\sim 0.02$ near the ends of the lobes. 
With $i = 50\degr$, we would have a modest $\beta = 0.25$ near the core,
which we will refer to as our `slow' model.
{\em In this case
the jet brightness asymmetry could not be due to beaming}, and so if
these associations are real we must abandon the conventional
interpretation of the Laing-Garrington effect, at least in Her~A.
Note that the arguments leading to the fast and slow models both 
constrain $\beta \cos i$; therefore any changes to the estimated inclination
$i$ would affect both speeds in the same way, with the inconsistency between
the two left unaltered.

If we retain $\beta\cos i \approx 0.5$ (fast model) as deduced from
beaming, corresponding features will differ by a factor of 3 in their
distance from the core, and the associations in
Table~\ref{correspondence} would be chance coincidences, or caused by
symmetries in the environment (which of course would be unaffected by
light travel effects). We have already noted one plausible example:
the transitions in the jets at a projected radius of $\approx 20$
arcsec seem to be associated with their entry into the bulbs.  In
turn, the distance of the inner edges of the bulbs from the core may
be controlled by the cluster core radius at 43 arcsec (see Paper I).
In this case the similar spacing between the rings in the eastern and
western lobes is puzzling, as the eastern ones should be spaced about
three times further apart; in other words, the western rings would
correspond to much more significant features in the jet than the rings
we have identified on the eastern side. As we discuss in
Section~\ref{disrupt}, this may be consistent with the more
clearly-defined structure of the western rings.

\subsection{Restarting jets?}
\label{restarting}

The spectral index map (Fig.~\ref{spixm}) shows that
the western jet and rings form a single coherent structure which is clearly 
distinct from the surrounding lobes; 
the geometry alone strongly suggests that these flat spectrum components
represent a renewed outburst of central activity within an old
lobe.  

Numerical simulations of this scenario have been made by
\citet{Wilson1984} and \citet{CB91}. They found that the 
hotspots expand and fade on their internal sound-crossing time
once the jet switches off. The lack of bright features at the 
outer edges of the western lobe of Her~A implies that enough time has passed 
since the last major outburst for this to occur, leaving the
lobe with a `relaxed' structure.  
The simulations show that 
the restarted jet is over-dense because its ambient medium consists of a
cocoon of rarefied and hot material from the original jet, which is under-dense
relative to the quiescent intergalactic medium. This implies a higher
advance speed for the restarted jet, but also a lower Mach number since
the new ambient temperature is much greater. A further 
implication is that the bow shock excited by the new jet is weak. 
This neatly explains why there is no true hotspot at the end of the western 
jet (i.e. near the bright arc Ah), because the new jet has not yet 
encountered the lobe surface. Over-dense jets should be terminated by 
a weak shock, and only a small amount of jet material will be reprocessed 
to form an `inner lobe' surrounding the end of the jet 
\citep*[c.f.][]{NWS1983}. In the limit of a very
low cocoon density, the advance speed will be equal to the jet flow
speed, and we assume this for simplicity in the following. 
\citet{Meier.etal1991} also concluded that the jets in Her~A are
heavy, but lacking the spectral index evidence they assume that the jet 
parameters must be extraordinary, whereas in our picture the jets are
fairly typical except for effects associated with their recent re-birth.

On the eastern side, although the jet is distinct from the surrounding
lobe over most of its length, there is a smooth change in spectral
index at the end. Here the most striking spectral division is between
the circular bulb, and the surrounding steep-spectrum bridge. 
This is consistent with the light-travel time delay across the source:
the new outburst is seen just before it reaches the end of the lobe on 
the western side, while on the eastern side it has already got there,
allowing fresh material from the jet to enter the outer part of 
the old lobe.  This material will be younger and so should
have a flatter spectrum than the material which was in the lobe before,
so the region of the lobe that it has reached will have a flatter
spectrum.  An obvious problem with this scenario is the absence of the
expected hotspot at the end of the eastern jet; this will be discussed
in Section~\ref{disrupt}.

We can estimate the timescales for the renewed outburst quite simply.
If $i \approx 50\degr$ and the jet speed is $0.8c$  (fast model), 
as suggested by beaming arguments, then
we see Ah (taking it to mark the end of the western jet) about 1.2~Myrs 
after the current outburst began. If  $v \approx 0.25c$ (slow model), as 
suggested by the apparent symmetries in Table~8, this becomes 3.8 Myr.
At the core, about 1.8 (4.5) Myr has passed for the fast (slow) model, 
because of the light-travel delay across the source, and at
the end of the east lobe 2.5 (5) Myrs has passed. 
The individual flares which cause the major rings would 
be separated by $\sim 250$ (800) kyr.  

If the lobes of Her~A are purely relativistic plasma their sound speed is
$c/\sqrt{3}$; relatively small amounts of mixing with the surrounding
intra-cluster medium (ICM) could reduce this. We then need $\ga 1$ Myr
for any hotspots associated with the previous outburst to expand to
equilibrium with the rest of the lobe.  For comparison, the
`spectral age' at the end of the western bulb is $\ga 13$~Myr, based on 
a minimum-energy magnetic field of $\approx 0.8$ nT and a fitted break 
frequency $\la 24$~GHz, both assuming 
the integrated spectral index of $\alpha = -1.0$.
Thus many flow-through times (or several, for the slow model) could have 
passed between the current outburst and the last time the lobes were actively
powered.

The eastern bulb may be a zone of turbulent mixing between 
the old and new
material; during this new outburst, the jet is unlikely to
have delivered sufficient material to completely fill such a large
region. In the `fast' model, we see the
end of the east lobe 2.5 Myr after the outburst starts, twice the time for 
material to flow all the way down the jet, so the contents of only one
jet length have been delivered to the lobe (even less, for the slow model).
Note that the western bulb has received no new material, except for 
the thin column containing the jets and rings, and yet its volume is about 
the same as the bulb on the eastern side. Thus the extra material 
delivered on the east should form only a small fraction of the lobe contents.
This is consistent with overall lifetime arguments:
we expect that Her~A is $\ga 11$ Myrs old (overall expansion
speeds are $< 0.1c$, \citealt{Scheuer1995}), and so in 1 or 2 Myrs the jet 
should deliver $\la 10$ per cent of the matter that fills the lobes, whereas 
the bulb occupies more than half the volume on each side. 

The time delay may also explain the the difference in width between
the two jets (c.f. Section~\ref{collim}): if the structure is
changing, we will see it at different times in its development.  The
response of a jet to a sudden increase in power has been studied by
\citet{Komissarov1994}. According to this study a significant increase
of the kinetic luminosity of the central engine results in the jets
becoming over-pressured and beginning to expand transversely. The
expansion is confined by ram pressure of the external gas at a
quasi-cylindrical shock around the jets, until the jets reach a
sufficient diameter that adiabatic expansion of the jet material
arriving from the AGN brings it into equilibrium with the ambient
pressure.  For a relativistic jet the time delay means that we would
see the jet at its narrowest at its most distant point, corresponding
to the initial low kinetic luminosity of the central source, and it
should widen systematically from there back towards the core (except
in the central $\la 7$ arcsec (20 kpc), where the jet expands from
the parsec scale to its quasi-equilibrium width), and continue to
expand on the other side away from the core until equilibrium is
reached. There is some evidence in Fig.~\ref{collim} that the western
jet narrows away from the core as this model predicts; evidently
equilibrium is established beyond $\approx 10$ arcsec east of the
core.

Spectral structure similar to Her~A's is found in two other DRAGNs.
\citet{Roettiger.etal1994} argued that 3C\,388
shows two distinct epochs of jet activity on the basis of a sharp spectral
index boundary between the bright inner lobes and surrounding `relic' lobes.
In this case the new outburst has reached a more mature stage, with a
well-defined hotspot at the end of the brighter jet. In 3C\,310, 
\citet{LPR86} 
found that the inner components B and D of \citet{Breugel.etal1984} 
had substantially flatter
spectra than the surrounding lobes. Van Breugel \& Fomalont already
drew attention to the similarity between the edge-brightened feature B 
of 3C\,310 and the western rings of Her~A; the spectral index structure
re-inforces this.  3C\,310 also contains a prominent (and complete) ring
in its southern lobe, but this is spectrally part of the `old' lobes. In this
case the young features extend only halfway to the edges of the lobes, 
so the new outburst is at an earlier stage than in Her~A.

All three DRAGNs are in relatively X-ray bright galaxy groups or
clusters \citep[Paper I;][]{Leahy.etal1999,Hardcastle.1999}, 
so the old lobes can be considered as examples
of `relaxed doubles' which are common in cluster centres.  Steep spectrum
DRAGNs are believed to occur in clusters because the high pressure 
environment prevents lobe expansion and associated adiabatic losses, 
giving time for substantial spectral losses to occur.  
The observation of renewed jets/inner lobes as spectral index features
in cluster DRAGNs may then be a selection effect: outside clusters, the old 
lobes could fade to invisibility before the next major outburst occurs.  
Other evidence for renewed outbursts in non-cluster DRAGNs is discussed
by \citet{LP92} and \citet{Schoenmakers.etal2000}.

\subsection{The disruption of the eastern jet}
\label{disrupt}

FR\,II DRAGNs are notable for the high collimation which allows them
to form very compact hotspots at the lobe ends.  However, we have seen
that the Her~A eastern jet seems to disrupt at knot E8; as a result it is
rather broad by the time it ends.  The detailed structure in the
decollimation region gives several clues to the the nature of the
catastrophe. Knots in jets are often attributed to internal shocks induced by
interaction with backflowing material in the `cocoon'
\citep{N82}; but these conical shocks do not much resemble what
is seen in Her~A, and are at least partially artefacts of the imposed
cylindrical symmetry.
Instead, knots E8, E9 and E11 suggest the sort of internal `working
surfaces' expected from time dependent speed variations in an over-dense jet,
which have been intensively studied in the context of jets from young
stars \citep[e.g.][]{SN1993,Cerqueira.etal.2001}. Such variations
could be caused partially, at least, by the central engine but may be
aided by local instabilities. Although the most detailed numerical
modelling has assumed radiative, and hence nearly iso-thermal, flows,
qualitatively similar structures occur in adiabatic, relativistic
flows, as we expect in Her~A \citep{KF1997}.
The disrupted structure at knots E9--E10 resembles the combination of
velocity and direction oscillations of the jet source, for jet beams
surrounded by hot, low density `cocoon' studied by \citet{RB1993}, and
also the complex structure of jets with well-developed
Kelvin-Helmholtz instabilities \citep[e.g.][]{Rosen.etal1999}.
Jet disruption is a characteristic feature of numerical
simulations of DRAGNs \citep{HN1990,Clarke1996} unless artificially
suppressed by enforced axial symmetry; to the point, in fact, where
the stability of real jets seems rather mysterious. 

An alternative mechanism for jet disruption is the passage of the
jet through an interface such as a shock in the ambient medium
\citep*{N88}, for instance caused by the impact of a galactic wind
on the surrounding medium. A serious drawback with this picture is that
high-resolution X-ray imaging has shown no sign of such shock fronts
around galaxies. Our own {\em ROSAT} data is probably too noisy to rule out
the idea in Her~A itself, but no sign of such a shock is seen in a
{\em Chandra} image kindly shown us by A. Wilson.  A more obvious
interface is the point of entry of the jets into the bulbs of the lobes,
if, as argued in Section~\ref{bridge}, the inner jets are directly confined
by the cluster core.  But this interface must be present in most FR\,II
DRAGNs, yet their jets almost always survive to form hotspots.

In any case, if the unusual structure of Her~A is caused by the recent 
restarting of its jets, this must explain {\em all\/} the abnormal features,
including the jet disruption, because it is very unlikely that one
object should be pathological in two independent ways.  
How, then, can we explain why a recently re-started jet should be
disrupted? Possibly, the lobe
material provides enough of an obstacle to
trigger instability, until a channel to the lobe surface is fully
established. But in our fast model, the eastern jet reached the end of 
its lobe about 1.2 Myr ago, long enough for a complete flow-through, 
so we would expect
the channel to be as stable by now as at any later time (this explanation
works better in our `slow' model, in which only a third of a 
flow-through has happened).
A more plausible scenario is that the AGN is far from a
perfect jet engine: the renewed outburst is not a sudden
return to a uniform flow, but instead fluctuates with a wide range of
timescales. Such an irregular flow would disrupt at a distance from
the engine set by the need for irregularities in the flow to form
shocks and/or breaks in the jet \citep[e.g.][]{RB1993}.

The idea of an irregular jet is supported by the structure on the
western side.  The bright features Eh and Ch, which are exactly
aligned with the inner jets, seem to correspond to very strong
internal shocks in the western jet -- they are many times brighter
than the assumed intervening sections (which are essentially
invisible).  The contrast is greater than for any knot in the eastern
jet.  In the fast jet picture, the western jet appears three times
slower than the eastern, and so represents three times the time span.
We would then expect to see more extreme internal events in the
western jet if the amplitude of an event is inversely correlated with
its probability of occurring; in other words, if irregularities in the
flow occur on successively longer intervals with successively larger
amplitudes. It seems that the output of the AGN splutters with a
roughly $1/f$ power spectrum.  It is worth emphasising that on this
interpretation, the features in the western lobe corresponding to most
of the eastern rings would appear as weak structure in the major
western rings; of course forshortened by a factor of three due to
light-travel effects.

The disruption of the eastern jet does not fully explain the lack of a
hotspot on that side. If the jet is over-pressured as
suggested by \citet{Meier.etal1991} then it should expand in a conical
morphology at its Mach angle and although the collimation beyond E8 is
poorer than for normal FR\,II jets, the implied Mach number would
still be ${\cal M} \sim 10$, so the jet should shock on contact with
the external medium. This suggests that at least the last section of
the jet, component F, is trans- or subsonic, and confined by the lobe
pressure.  We argue below that in fact this is the only subsonic
portion, that is, the flow up to knot E13 is supersonic.

In fact it might not be easy for the jet to decelerate while in the lobe.
In the standard model for FR\,Is \citep{Bick84},
jets decelerate by sharing their momentum
with entrained ambient medium. But the lobe material should have a low density 
compared to the jet, which will make this mechanism rather ineffective.
This leaves the impact on the lobe surface (that is, on the true intracluster
medium) as the main deceleration mechanism. Nevertheless, if the jet material
has been pre-heated and decollimated by internal shocks, this will not result
in a compact hotspot.

\subsection{The nature of the rings}
\label{rings}

As noted by \citet{Meier.etal1991},
the key to understanding the western rings in Her~A is the 
observation that similar features exist in the eastern lobe. These eastern
rings are closely associated with the eastern jet, surrounding it, 
but are clearly separated from the ridge-line marking the jet itself.  
We have already noted the evidence for faint compact structures within the 
western rings which may be part of a continuation of the counter-jet, 
so the apparent difference between the two sides is mostly just due to
the fact that the counter-jet is much dimmer than the main jet; additional
differences arise from the light-travel time delay, of course.
The structure of the rings
suggests two hypotheses: either the rings are a system of shocks induced
by the jets in the surrounding lobe plasma, or the rings form an `inner
lobe' around the jets, consisting of material deposited by the new jets,
and separated from the old lobes by a contact discontinuity. In either
case, the material in the rings is not moving with the bulk velocity of
the jets. This is particularly important in our `fast' model as otherwise
we would expect the same brightness asymmetry between the rings on the
two sides as between the jets, whereas in fact the two sets of rings are
similar in brightness.  It will become apparent that these two models are
not as distinct as first appears, and a combination of the two may be
closer to the truth.

\subsubsection{The rings as shocks}
The best argument for the shock model is that the shape of the rings,
especially around the eastern jet, strongly resembles the `side shocks'
known from stellar jet, that is, shocks
propagating through the ambient material, 
caused by the passage of bumps on the jet surface (except ring G which seems
to be a bow shock ahead of knot E13). 
There is a strong resemblance
to the HH\,47 stellar jet \citep{Heathcote.etal1996}, where shocks in
the surrounding neutral wind can be clearly identified by their 
H$\alpha$ emission.  Because of radiative cooling, HH\,47 contains material
with a wide range of densities but the side shocks are attached to dense
clumps in the jet identifiable by strong [S{\sc ii}] emission; thus like
Her~A it is effectively a dense jet.
Jet surfaces can become irregular 
through variations in both the direction and speed of the jet, 
and also from the growth of instabilities \citep{Heathcote.etal1996}. 
All of these are certainly present in HH\,47,  
and we have seen that the same may be true in Her~A.

The relatively flat spectrum of the rings is not easy to explain in
this model. The low-frequency
integrated spectrum of Her~A is very close to $\alpha_{\rm Low} = -1.0$
\citep{Kuhr.etal1981}, and this is probably dominated by the bulbs.
As we have seen (Table~\ref{spix}), the lobe spectra steepen at GHz
frequencies.  If the rings are shocks, adiabatic compression would shift 
the curved portion of the spectrum to higher frequencies, so that between 
a fixed pair of frequencies the spectral index would flatten, to 
limiting value set by $\alpha_{\rm Low}$.  But the rings have
a flatter spectrum than this, so we have to invoke particle acceleration,
presumably by the shock-Fermi mechanism.  This may not seem surprising,
as shocks are believed to be the sites of particle acceleration in most
radio sources. But it is then curious that the
steepest-spectrum ring is Ah, which should mark the bow shock ahead of
the Doppler-dimmed western jet, and therefore be the strongest shock of
all (the internal bow shock ahead of the eastern jet having disappeared
when it reached the end of the lobe). In fact, Ah is the {\em only}
ring whose spectrum requires only adiabatic compression of the lobe
material. Another problem for the shock model is the similarity between
ring and jet spectra, which would have to be a coincidence.
However, these are similar
mainly in contrast to the extremely steep spectra in the undisturbed old
lobes; there is a systematic spectral difference of about 0.1 and 0.2
between jet and  rings in the eastern and western lobes respectively 
(Table~\ref{tomo}), although it happens that $\alpha(1.5,4.8)$ in 
the western rings is close to that of the eastern jet.  

In fact the neat division into old lobe and new jet material is an
oversimplification in at least one place. We argued in Section~\ref{disrupt}
that the material in the fan component F at the end of the eastern jet is 
at most trans-sonic. In contrast, the region up to at least 
E12, where side shocks are clearly seen, must (on this interpretation)
be highly supersonic, because side shocks imply that the jet is 
supersonic with respect to the lobe,
which should be hotter and therefore have a higher 
sound speed than the jet. 
We would therefore expect a shock in the fan material
as the faster jet behind ploughs into it, and the ring G seems a plausible
candidate for this.  Some of the other rings may also be partly interactions
with slower sections of the jet, rather than the undisturbed old lobe.

In side shocks, the angle of the shock front to the jet flow direction 
should tend to the Mach angle 
at large distances from the jet axis, providing that the bump on the jet 
which causes the side shock persists for long enough
for the shock to reach this regime.
In HH\,47 the shocks become almost parallel to the jet, so the motion must be 
hypersonic in the surrounding medium, and this is confirmed by measurement 
of the shock speed and ambient temperature.  
The structure in Her~A is similar, although we note that if the jet is 
relativistic the apparent angle of the shocks will be affected by time-delay 
effects, giving an apparent Mach angle
\begin{equation}
\theta_{\rm app}  = 
\arcsin\left({\gamma_s \beta_s (1 \pm \beta \cos i) \over \beta \sin i}\right),
\end{equation}
where $\gamma_s \beta_s$ refers to the sound speed, and the minus sign applies
in the approaching jet.
If the lobes are pure relativistic plasma we have $c_s = c/\sqrt{3}$
and hence $\gamma_s \beta_s = 1/\sqrt{2}$. For our fast model ($\beta = 0.8$
and $i \approx 50\degr$) we expect $\theta_{\rm app} \approx
36\degr$ in the approaching lobe, which is much larger than observed in 
rings I and J; while in the receding lobe we should not see a Mach cone 
at all. Similarly, in the slow model the jets would be subsonic if the lobe
plasma was relativistic.  Thus to make the shock model work we have to assume
that enough gas from the ICM has been mixed into the lobes to
reduce the sound speed.  This need not be much: if the pressure is
dominated by relativistic plasma, the sound speed is $\sqrt{4 P/3 \rho}$.
If we take $P \approx 13$ pPa (pressure balance with the ICM), then
to reduce the sound speed by an order of magnitude below the relativistic 
limit we would need a proton number density of $n \sim 16$~m$^{-3}$, still
far below the density in the ICM. 

If we do consider the ring opening angle to be related to the Mach
angle, then there is some evidence for deceleration along the eastern
jet: the apparent shocks in component H show a significantly larger opening
angle than the ones in components I and J. Ring G, in contrast is tightly 
wrapped around knot E13. This
is consistent with G being a shock in the jet as E13
runs into component F, because the jet material should be significantly 
denser than the lobe material and hence has a lower sound speed.

The argument for entrained material in the lobes would be circumvented if 
the jet knots causing the shocks were
expanding transversely; in the limit that the knot expands rapidly, 
the shock will form an ellipse around the expansion centre. 
Something like this {\em must} occur for rings E and C, if they are shocks 
at all, as they clearly narrow from their maximum width towards the core, with
C closing completely on the upstream side.  
The three mechanisms suggested in
Section~\ref{disrupt} for jet knots (in brief, shocks caused by
pulsed jets, shocks caused by variations in the jet direction, and
shocks developed by Kelvin-Helmholtz instabilities) would certainly
provide some sideways expansion; in effect these provide a
fluid-dynamical explanation for the sequence of `explosions'
postulated by \citet{Mason.etal1988} to explain the rings. Again, the arc
Ah is the hardest case: the jet end would have to be expanding sideways
at about the same speed as it moves forwards to create such a circular
shock, and this would have to be transient as otherwise the jet opening
angle would be a full radian!

\citet{SBS02} propose an alternative model in which the rings, rather than 
being partial shells seen in projection are annular
shocks such as those surrounding jets in axisymmetric numerical simulations.
We find their model unconvincing because it
does not explain why the rings are brightest on the side more distant from the 
core, and it predicts that the long axis of the rings should be orthogonal 
to the jets, whereas most are elongated parallel to the jets. As in our
shock model,  the similarity between jet and ring spectral indices would 
have to be a coincidence.

\subsubsection{The rings as inner lobes}

The alternative hypothesis that the rings are the new inner lobe is suggested
mainly by their similarity in spectral index to the jets: 
of course this removes
all the problems of the shock model in this regard.  Because the jet is
over-dense, far less material is fed into the lobe than in the usual case
where a light jet impacts directly onto the intra-cluster medium; therefore
the inner lobe is narrow, not much wider than the jet itself. 

If the system of rings as a whole constitutes a pair of inner lobes, what
are the individual rings?  The evidence of the eastern lobe still points
to a close connection between rings and at least some of the knots in the
jets.  The references discussed in Section~\ref{disrupt} describe how
internal working surfaces in the jet will eject material transversely. 
In a narrow inner lobe, this material will impact on the surrounding contact 
discontinuity, causing a bulge in the lobe surface around each knot.
In fact, these interactions will be mediated by side shocks very much as
discussed in the previous sub-section, although this time they will be
propagating through the inner rather than the outer lobes.

In the inner lobe model, 
the arc Ah is the material between the jet shock and the 
contact discontinuity. 
It is visible because the jet decelerates at the jet shock and so is no
longer anti-beamed, and also because the pressure is higher in this
region (but not much higher, because the bow shock and jet shocks are
weak).

The steep-spectrum filament ahead of Ah (Fig.~\ref{tomomap})
seems an excellent candidate for lobe material swept up ahead
of the new outburst, perhaps bounded by a weak shock. Because the shock
is weak, we would not expect significant particle acceleration and only
mild compression: the steep spectrum is then explicable, especially
if this material has been carried out from the centre of the source, 
where the spectrum is very steep.
Similarly, the apparent way that the faint intensity filaments in the western
bulb arc around the ABC complex supports this hypothesis: on the shock
model this part of the lobe should not yet be aware of jets.

The major drawback with the inner lobe interpretation is the strong 
rim-brightening of the rings: normal lobes do not usually show this feature
(although in fact the western bulb of Her~A is somewhat edge brightened, as
shown in Fig.~\ref{xtoti}).
At least in part this may be due to the narrowness of the inner lobe:
the centre of the lobe is taken up by the jet, which is invisible due to
Doppler anti-beaming, and so the lobe appears hollow. This is not the
whole story, because we still see offset rings on the east side; but
numerical models suggest that the jets should be bounded by an ultra-hot
and low density sheath, which will be more apparent when the lobe is narrow.

\section{Conclusions}
\label{conclusion}

Our observations of the powerful radio galaxy Her~A have
shown a strong Laing-Garrington effect: the jet side depolarizes less
than the counter-jet side with increasing wavelength.  The fact that
depolarization is wavelength dependent is a characteristic of Faraday
rotation \citep{Burn1966,Laing1984}, which will be discussed in more
detail in Paper III.

We also find that the less depolarized lobe has a flatter spectrum,
and that this is not simply due to the presence of the prominent (and possibly
beamed) jet.  This is in agreement with the trend discovered by 
\citet{Liu.etal1991}.

After correcting for Faraday rotation the projected magnetic
field closely follows the edge of the source, the jets, and the
ring-like structures in the lobes; the field pattern in the two lobes
is broadly similar.

We have discovered a remarkable structure in spectral index, qualitatively
similar on the two sides, which strongly suggests the recent ejection
of high-brightness, relatively flat spectrum material into much steeper
spectrum lobes. 

We have argued that all the peculiar features of Her~A can be understood
if the outflow from the AGN effectively ceased for $\ge 1$ Myr,
restarted some 2-5 Myr ago (depending on the speed of the jets), and has
since fluctuated with a roughly $1/f$ power spectrum. Of course, the overall
fluctuations on timescales of several Myr might well be part of the same
spectrum; the present episode may not be the first major re-start. 
At the other end of the fluctuation frequency spectrum, it may be
significant that the compact ($\sim 10$-pc scale) core lacks
bright sub-pc structure (since there is no flat-spectrum component). 
This suggests substantial variability on timescales of decades, with the 
outflow currently low or off. 

Because re-starting jets propagate rapidly through the old lobes, Her~A
will only show its present peculiar structure for a small fraction of its
life: rather than being a highly pathological AGN, we suspect that it is
a fairly typical powerful DRAGN, caught at a very atypical phase in its
development. 

We have shown that a consistent model can be found in which the jets are
fast ($v \approx 0.8c$) and the systematic asymmetries between the two jets
are accounted for by relativistic beaming, together with other effects 
related to the light-travel delay across the source in the presence of
fluctuating, but symmetric, outflow from the AGN. This model requires us
to abandon some of the apparent morphological symmetries between the two
lobes as merely accidental; but we have argued that these symmetries are
not very impressive.  This model is consistent with the conventional
interpretation of the Laing-Garrington effect as strong evidence for
relativistic flow in large-scale jets, and can also explain the Liu-Pooley
asymmetry, at least for Her~A.

We have also considered a  model in which the
apparent morphological symmetries reflect a series of pulses in `slow' 
($v \sim 0.25c$) jets. On this model, though, the main jet would be 
intrinsically brighter than the counter-jet and so the relativistic
explanation of the Laing-Garrington effect would not apply, at least
in this particular object. Both fast and slow models require fluctuating
outflow from the AGN on timescales much shorter than the overall re-start.
In general the slow model is less well constrained.

We considered two models for the famous `rings' (which are found around
both jets): that they represent a system of shocks in the old lobes
excited by the renewed jets, or that the represent the surface of an
`inner lobe' surrounding the jets.  When their consequences are followed
through, both models imply that the rings are related to shock systems
which at least partially involve interaction between components in the
new outflow; thus a hybrid of these simple models may be the most plausible
option. On any 
interpretation, the material in the old lobe must significantly obstruct
the new jets: although for simplicity our kinematic models assumed
no deceleration, this must be a rather rough approximation. Numerical
simulations of re-starting relativistic jets could shed much light
on the structures we have observed.

Our interpretation of Her~A as showing restarting jets 
is entirely consistent with a simple interpretation of the
spectral structure in terms of spectral ageing, which suggests that
the jets and rings (on either interpretation) 
with their flatter and less curved spectra are significantly
younger than the lobes \citep[c.f.][]{Leahy1991}. Of course, 
a detailed spectral ageing analysis should also take into account the 
effects of adiabatic expansion on the spectrum.  We are planning such a study, 
incorporating new high resolution data at 330 and 74 MHz
\citep[and in preparation]{Kassim.etal1993} 
to better constrain the spectral fits.

\section*{Acknowledgments}
We thank Simon Garrington and John Dreher for useful discussions.
John Dreher also gave us an unpublished 5-GHz image which inspired some of 
our interpretation prior to the full reduction of the images presented here,
and Andrew Wilson kindly shared his {\em Chandra} image in advance of 
publication. We thank Rick Perley for advice and help with the observations, 
and Simon Garrington for providing the calibrated D-configuration data
at 5 GHz. 

Nectaria Gizani would like to acknowledge PPARC for funding her fees
for three years to carry out her PhD work at Jodrell Bank at the
University of Manchester. Also the BFWG Charitable Foundation, which
supports women in science and kindly gave her an additional
grant. This work was sponsored by the Funda\c{c}\~{a}o para a
Ci\^{e}ncia e a Tecnologia, Portugal, under the contract number
PRAXIS XXI/BPD/18860/98.

The National Radio Astronomy Observatory is a facility of the 
National Science Foundation operated under cooperative agreement by 
Associated Universities, Inc. 
This research has made use of the NASA/IPAC Extragalactic Database
(NED) which is operated by the Jet Propulsion Laboratory, Caltech,
under contract with the National Aeronautics and Apace Administration.
This research has also made use of NASA's Astrophysics Data System.


\bibliography{../3c348}

\end{document}